\documentclass[%
 aip,
 jmp,%
 amsmath,amssymb,
 reprint,%
]{revtex4-2}
\usepackage{xcolor}
\usepackage{subcaption}
\usepackage{hyperref}%
\usepackage{import}
\usepackage{tabularx}
\usepackage{verbatim}
\usepackage{booktabs}
\usepackage{mathtools,tikz,caption}
\usepackage{rotating, graphicx}
\usepackage{float}
\usepackage{caption}

\DeclareRobustCommand\sampleline[1]{%
  \tikz\draw[#1] (0,0) (0,\the\dimexpr\fontdimen22\textfont2\relax)
  -- (2em,\the\dimexpr\fontdimen22\textfont2\relax);%
}
\usepackage{graphicx}
\usepackage{dcolumn}
\usepackage{bm}

\begin{document}

\preprint{AIP/123-QED}

\title{Compressible Boundary Layer Velocity Transformation Based on a Generalized Form of the Total Stress}

\author{Hanju Lee}
\author{Pino M. Mart\'{i}n}%
 \email{mpmartin@umd.edu}
\affiliation{ 
Department of Aerospace Engineering, University of Maryland, College Park, MD 20742, USA}%

\author{Owen J.H. Williams}
\affiliation{William E. Boeing Department of Aeronautics and Astronautics, University of Washington, Seattle, WA 98195, USA}%

\date{\today}

\begin{abstract}
The effects of density and viscosity fluctuations on the total stress balance are identified and used to create a new mean velocity transformation for compressible boundary layers. This work is enabled by an extensive database of direct numerical simulations that incorporate wall-cooling, semi-local Reynolds numbers ranging from 800 to 34000, and Mach numbers up to 12. The role, significance and physical mechanisms connecting density and viscosity fluctuations to the momentum balance and to the viscous, turbulent and total stresses are presented,  allowing the creation of generalized formulations. We identify the significant properties that thus-far have been neglected in the derivation of velocity transformations: (1) the Mach-invariance of the near-wall momentum balance for the generalized total stress, and (2) the Mach-invariance of the relative contributions from the generalized viscous and Reynolds stresses to the total stress.  The proposed velocity transformation integrates both properties into a single transformation equation and successfully demonstrates a collapsing of all currently considered compressible cases onto the incompressible law of the wall, within the bounds of reported slope and intercept for incompressible data. Based on the physics embedded in the two scaling properties, the success of the newly proposed transformation is attributed to considering the effects of the viscous stress and turbulent stresses as well as mean and fluctuating density  viscosity in a single transformation form.

\end{abstract}

\maketitle

\section{Introduction}
Hypersonic flow is an area of interest that has received much attention by the fluid mechanics community in recent years as we move towards space travel and hypersonic civilian transport vehicles. In particular, a mean velocity transformation (MVT) for turbulent boundary layer (CTBL) flow that accounts for variations in thermodynamic variables and compressiblity has been sought.  The main driver behind such efforts was the suggestion by Morkovin in 1962 ``that for moderate Mach numbers, the essential dynamics of these shear flows will follow the incompressible pattern.'' \cite{SmitsDussauge} Based on this hypothesis, it has been hoped that a correct accounting for variation of thermodynamic properties will restore the incompressible law of the wall for wall-bounded compressible flow.\\
\indent Over the past few decades, various forms of MVTs for CTBLs have been proposed.\cite{VD,TL,ZhangHussain,Brun-Boiarciuc-Haberkorn-08,wu-Hussain-She-2017,Patel16}, from the pioneering work of Van Driest~\cite{VD}, to the recently proposed viscous stress based transformation by Trettel et al.~\cite{TL} (TL), each have  had their own limitations. The Van Driest (VD) transformation has shown success in scaling adiabatic CTBLs \cite{guarini_moser_shariff_wray_2000,LianMartinMach,lagha-kim-zhong-2011,pirozooli-gatski-2004} with reported weakness in scaling of iso-thermal cases leading to increases in the log-law intercept with increasing heat transfer~\cite{coleman_kim_moser_1995,foysi_sarkar_friedrich_2004,Patel16,TL,ZhangLian2018}. The most successful MVT that accounts for the near-wall viscosity gradient was independently developed by Trettel et al.~\cite{TL} and Patel et al.~\cite{Patel16}. However, despite their initial success in isothermal CTBL cases \cite{modesti-pirozzoli-2016,Yao-Hussain-2020} with low semi-local Reynolds, $Re^*$, the scaling remains unsuccessful for cases with increasing $Re^*$ where multiple studies\cite{GriffinMoin,ZhangLian2018} report a large scatter in the log-layer intercept and the slope for such cases. In this paper, the semi-local friction Reynolds number is defined as,  $Re^*=Re_{\tau}\sqrt{\overline{\rho}_e/\overline{\rho}_w}/(\overline{\mu}_e/\overline{\mu}_w)$, $u_{\tau}=\sqrt{\tau_w/\overline{\rho}_w}$ is the friction-velocity, $\delta$ is the boundary layer thickness and $\tau_w = \overline{\mu {\partial u}/{\partial z}}|_w$ is the wall shear stress. The fluid density and viscosity are denoted  by $\rho$ and $\mu$. The variables $u,v,w$ denote velocities in the streamwise, spanwise and wall-normal directions, $x,y,z$, respectively. The  conventional friction Reynolds number based on wall conditions is defined  $Re_{\tau}=\overline{\rho}_w u_{\tau}\delta/\overline{\mu}_w$. The overbar denotes time averaging. A single prime will be used to denote a fluctuation from the Reynolds-averaged mean velocity, and the subscripts 'w' and 'e' will denote wall or boundary layer edge quantities.\\
\indent Recently, Griffin, Fu \& Moin~\cite{GriffinMoin} (GFM) developed a total stress based MVT that employs a combination of the viscous stress transformation of Trettel et al.~\cite{TL} and the quasi-equilibrium assumption based transformation of Zhang et al.~\cite{ZhangHussain}. They employ a total stress-based functional form to combine these two transformations, such that each is  applied in their region of applicability; Trettel et al.~\cite{TL} in the viscous layer and Zhang et al.~\cite{ZhangHussain} further from the wall. In their transformation, the quasi-equilibrium model is extended to employ the semi-local wall-normal coordinate, $z^*=zu_{\tau}\sqrt{\overline{\rho}_w/\overline{\rho}}/(\overline{\mu}/\overline{\rho})$, which was proposed by Huang et al.~\cite{HuangColeman95} and has shown to be effective in collapsing turbulence statistics in CTBLs within the near-wall region \cite{TL,ZhangLian2018,LianMartinEnthalpy,LianMartinMach,GriffinMoin,patel15,Patel16,coleman_kim_moser_1995}. Initial explorations of  the mean velocity profiles transformed by the GFM approach have shown promising collapse and improvement over earlier MVTs.  However there are questions about the breadth of applicability of the quasi-equilibrium hypothesis on which the GFM transformation is partially based. A follow up paper by Bai et al.~\cite{BaiGriffinFu} extended this exploration for  a broad range of cases, including high enthalpy turbulent boundary layers, flows at supercritical pressure and boundary layers with pressure gradients, with mixed results, suggesting additional physics must be incorporated under these conditions.  The intercept and the slope of GMF transformation as well as the quasi-equilibrium assumption are further examined in the present paper for a broader set of semi-local Reynolds numbers.\\
\indent A few studies have derived MVTs from the momentum equation perspective and this approach deserves further attention. The transformation by Zhang et al.~\cite{ZhangHussain} is derived from the turbulent kinetic energy equation and Wu et al.~\cite{wu-Hussain-She-2017} examined the total stress from the momentum equation perspective to derive their MVT, for example. The latter study, however, relies on Prandtl's mixing length hypothesis \cite{Prandtl} for the prediction of the Reynolds stress and requires \textit{a priori} information of the onset locations for the buffer and log layers. Such requirements make the use of their transformation difficult to apply in practice. While relying on the Mach-invariance of the total stress in the momentum equation may be more practical, data show Mach dependence and the breakdown of the near wall momentum balance for CTBLs, as classically described.  Namely, for high $Re^*$ number and high Mach number, the turbulent stress is greater than the wall shear stress under  Reynolds averaging, albeit slightly \cite{LianMartinEnthalpy,LianMartinMach,ZhangLian2018,lagha-kim-zhong-2011}.  While density fluctuations can be accounted for by employing Favre averaging, this result suggests viscosity fluctuations might need to be considered to appropriately achieve Mach-invariance of the total stress and to derive MVT from the momentum equation perspective.\\
\indent The current paper makes use of the CRoCCo CTBL database, which includes a wide range of $Re^*$, approximately spanning from 800 to 34000, and is used in the present study to scrutinize the effects of Mach and Reynolds numbers on CTBL flows. In the present study, these effects are examined in the context of the thin shear layer (TSL) momentum approximation, which we modify from its classical form to represent the data across the parameter space. After the relative importance of density and viscosity fluctuation effects is analyzed, we integrate the new findings into a mean velocity transformation that enables the collapse of data across a wide range of CTBL conditions.

\section{Simulation Details and CTBL Database}
\subsection{Flow Conditions}
The direct numerical simulation database used for this study is summarized in Table~\ref{Compressible DNS Database}. All simulations employ low-enthalpy, non-reacting conditions typical of ground test facilities.  The working fluid is  callorically perfect air for all cases except for M10T3 which utilizes callorically perfect Nitrogen($N_2$). The boundary layer edge Mach number, $M_e$ ranges from 3 to 12 to highlight the Mach number effects. The semi-local friction Reynolds number, $Re^*$, ranges from 800 to 34000, approximately, to highlight Reynolds number effects in compressible turbulent boundary layer flow. While a number of cases are adiabatic, the $\overline{T}_w/\overline{T}_r$ ratio, as well as the wall heat transfer rate, $B_q=q_w/(\overline{\rho}_wC_pu_{\tau}\overline{T}_w)$, range from 0.2 to 1.0 and from 0 to -0.17, respectively, where $q$ is the surface heat flux, $C_p$ is the heat capacity at constant pressure and $\overline{T}$ is the mean temperature. The wall temperature $\overline{T}_w$ is given as a fraction of the adiabatic recovery temperature $\overline{T}_r = \overline{T}_e(1+0.9 M_e^2 (\gamma - 1)/2)$, where $\gamma$ is the ratio of specific heats. 
Several Reynolds numbers are provided, including $Re_\theta \equiv \overline{\rho}_e \overline{u}_e \theta/\overline{\mu}_e$ where $\theta$ is the compressible momentum thickness and $\overline{\rho}_e$, $\overline{u}_e$, and $\overline{\mu}_e$ are the boundary layer edge density, velocity, and dynamic viscosity respectively. A second momentum thickness Reynolds number, $Re_{\delta 2}$ is defined as $\overline{\rho}_e \overline{u}_e \theta/\overline{\mu}_w$, following convention. The friction Reynolds number at wall conditions, $Re_\tau$ varies between 475 and 825. All values of Reynolds number listed in Table~\ref{Compressible DNS Database} are measured at the outlet plane of the computational domain. The locations of the data collection outlet plane $x_o/\delta$ are provided in table \ref{tab:02} where $\delta$ is the thickness of the boundary layer at 99\% of the freestream velocity measured at the outlet plane. 
\subsection{Computational domain and simulation set-up}
\begin{figure}
\centering
\begin{subfigure}{.475\textwidth}
  \includegraphics[width=\linewidth]{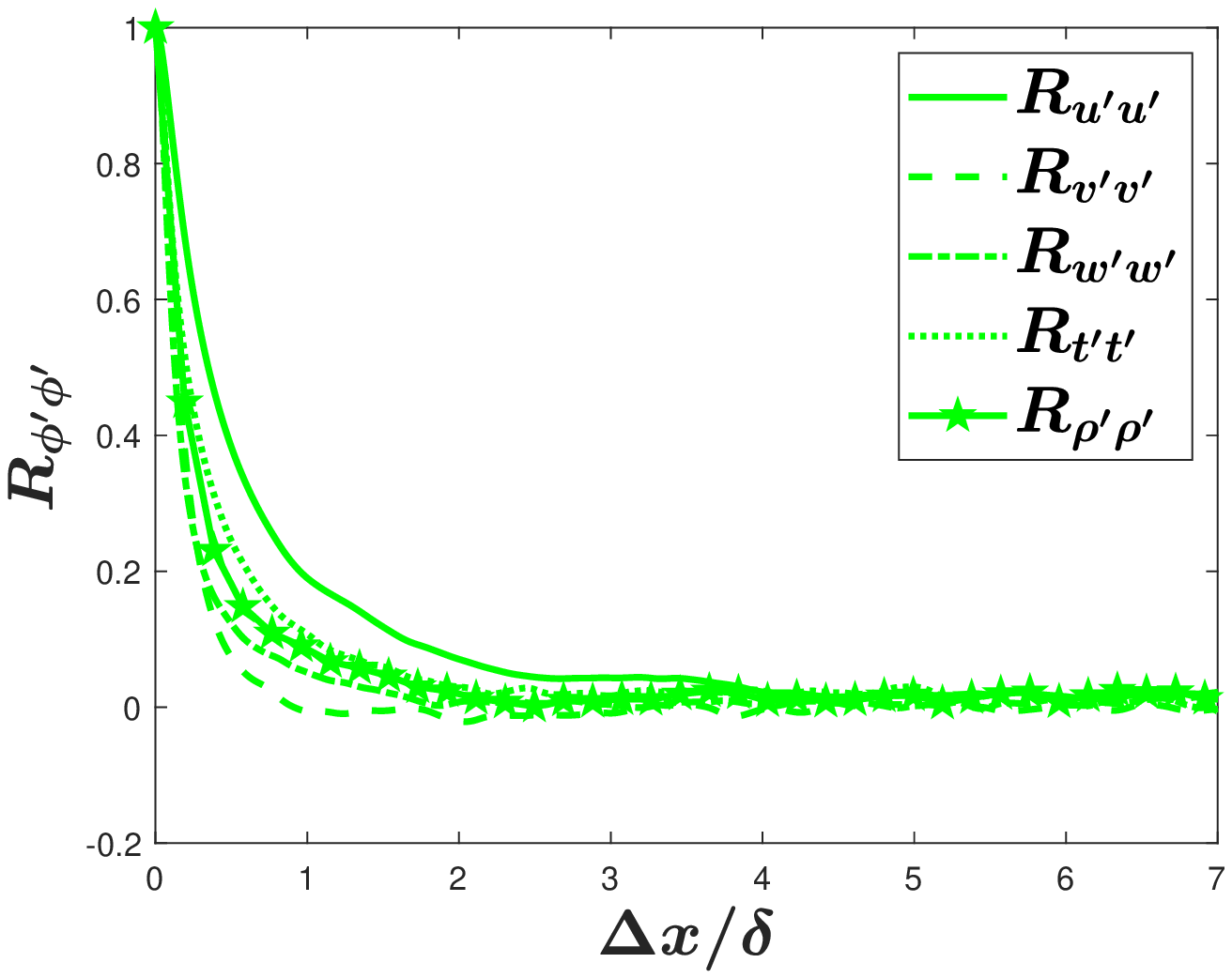}
  \caption{}
  \label{fig:M5T1_Rx_zdelta01}
\end{subfigure}
\begin{subfigure}{.475\textwidth}
  \includegraphics[width=\linewidth]{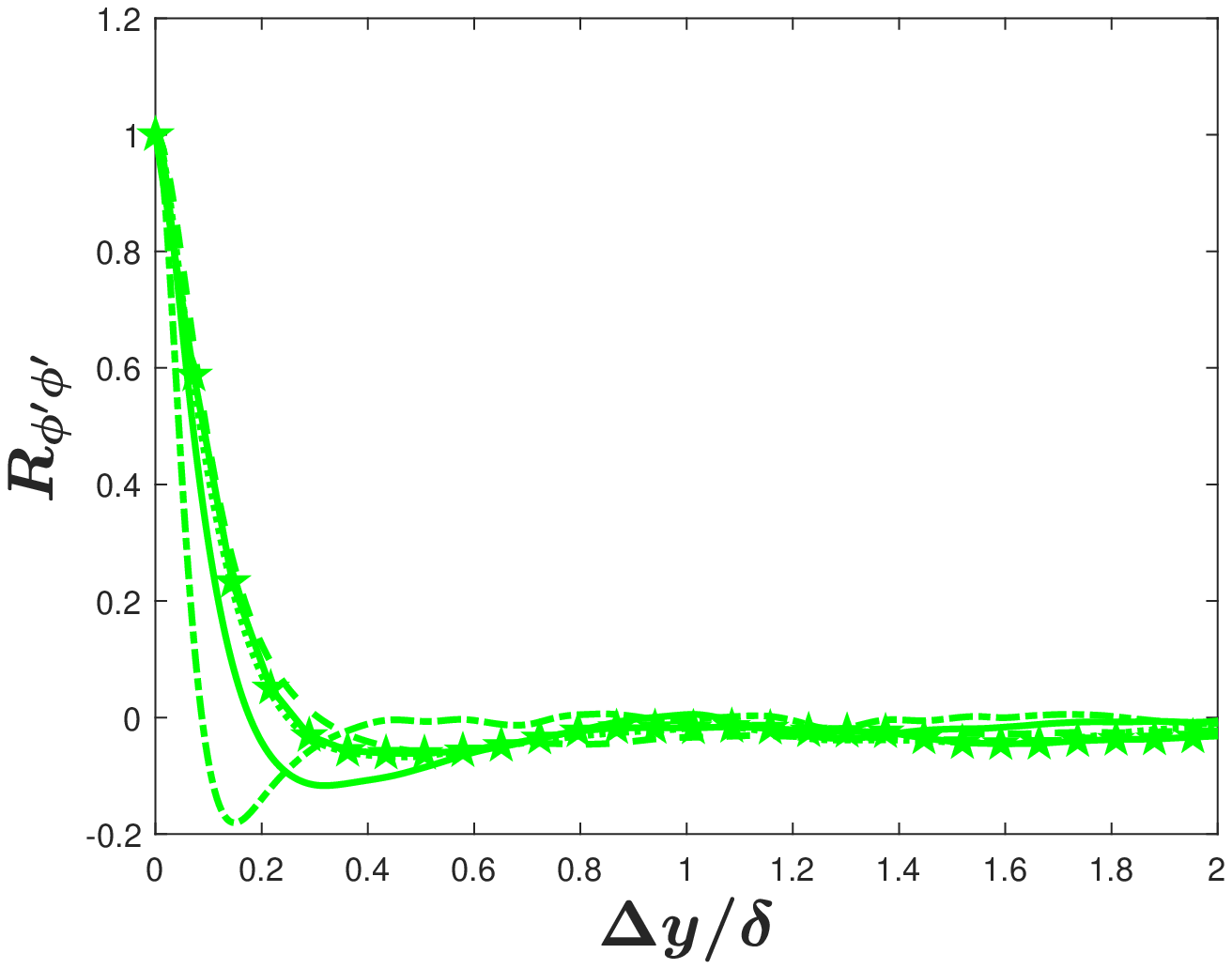}
  \caption{}
  \label{fig:M5T1_Ry_zdelta01}
\end{subfigure} 

\begin{subfigure}{.475\textwidth}
  \includegraphics[width=\linewidth]{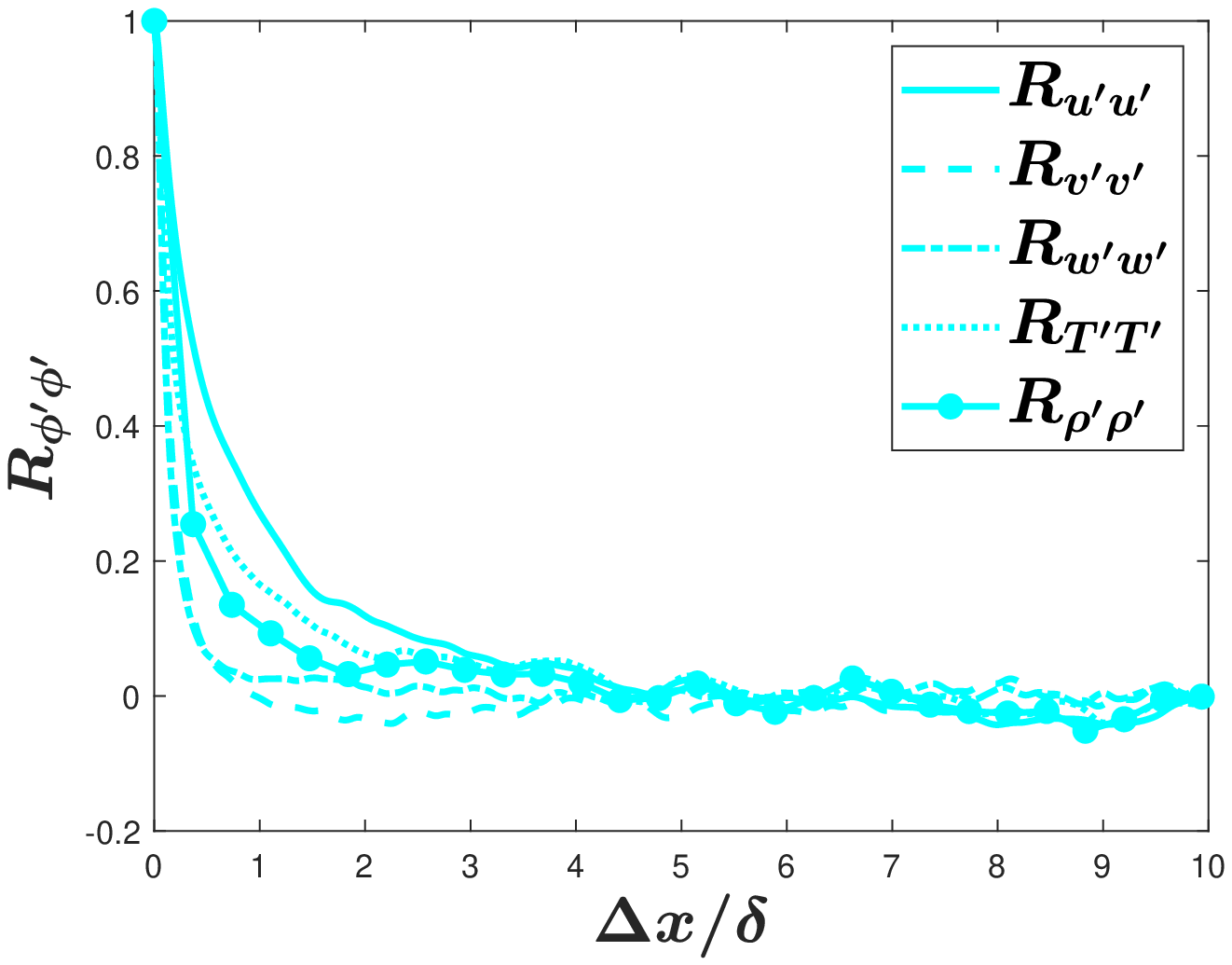}
  \caption{}
  \label{fig:M12T5_Rx_zdelta01}
\end{subfigure}
\begin{subfigure}{.475\textwidth}
  \includegraphics[width=\linewidth]{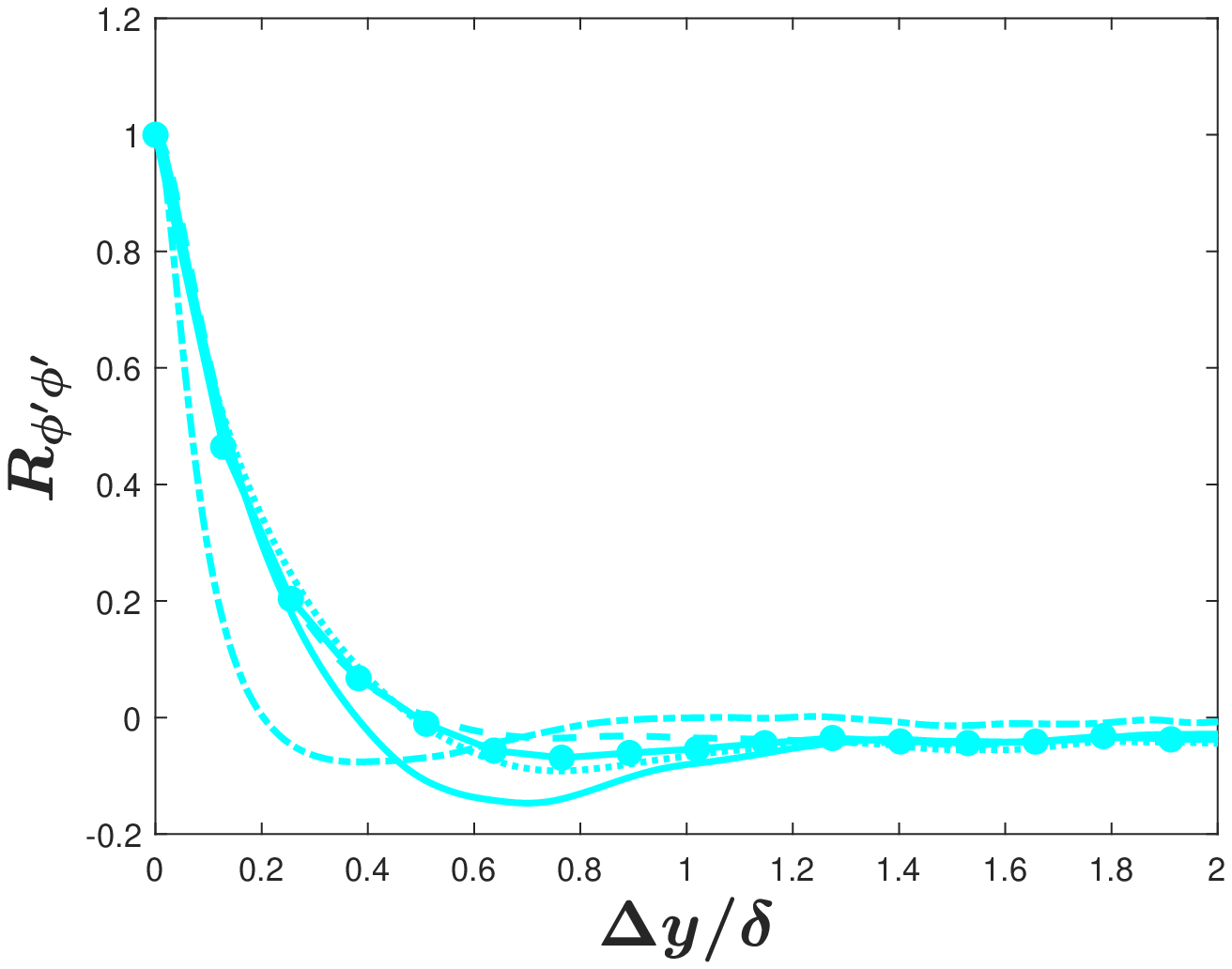}
  \caption{}
  \label{fig:M12T5_Ry_zdelta01}
\end{subfigure}
\caption{Two point correlation coefficient, $R_{\phi'\phi'}$, for streamwise, spanwise and wall-normal velocity as well as temperature and density components at $z/\delta = 0.2$. (A) Streamwise M5T1 (B) Spanwise M5T1 (C) Streamwise M12T5 (D) Spanwise M12T5}
\label{fig:Correlation}
\end{figure}

The governing equations, numerical methods, boundary conditions and initialization procedures used to create the high-fidelity CRoCCo Lab database for this study are documented and have been verified in previous studies~\cite{Martin07,XuMartin04,TaylorMartin06,TaylorMartin07}.\\
\indent The computational domain size, grid resolution, and simulation duration of the datasets are provided in Table~\ref{tab:02}. The outer dimensions of the computational boxes are given in units of $\delta$ measured at the outlet plane. All runs use spanwise periodicity. The domain width varies among the runs but ranges between 5 and $10\delta$.  All cases use the recycle/rescale method of Xu \& Martin\cite{XuMartin04} to assign the inflow boundary conditions. Large-scale structures are artificially introduced at the inlet due to the recycle/rescaling method, requiring domains that are long enough for the flow to become decorrelated with from the inflow.  For most cases, the domain was approximately 20 to 30 times the outlet boundary layer thickness. For M7T5-L and M12T5-L an extended domain length of approximately 40 $\delta$ was employed in two stages, first from $x=0$ to approximately 20$\delta$ in M7T5* and M12T5* (which we refer to as auxiliary cases), and then from about roughly 20$\delta$ to 40$\delta$ in M7T5-L and M12T5-L. To assess the adequacy of the domain size, a two point correlation coefficient of the streamwise, spanwise and normal velocity components, as well as the temperature and density at $z/\delta = 0.2$, is plotted for both M5T1 and M12T5 which are the cases with the lowest and the highest  $Re^*$, respectively. The streamwise and spanwise correlation coefficients in figure \ref{fig:Correlation} drop to near 0 at locations with a large enough separation, suggesting that the domains are sufficiently long. Similar results were observed for other cases.\\
\indent Grid resolutions are listed in Table~\ref{tab:02} relative to the wall-referenced inner viscous length scale $z_\tau = \overline{\mu}_w/\overline{\rho}_w u_\tau$ as indicated by the `$+$' superscript. The computational grids are made with constant spacing in the streamwise and spanwise directions ($\Delta x^+$ and $\Delta y^+$). Geometric stretching is used in the wall-normal direction where $z_k^+ = z_2^+(\chi^{k-1}-1)/(\chi - 1)$ and $k$ indicates the wall-normal grid index so that the finest resolution is at the wall surface. The first grid point away from the wall is $z_2^+$ and the factor $\chi$ determines the rate of stretching. The grid resolution of the DNS runs has been shown to be sufficient for the current DNS computational method and at the given flow conditions in previous publications. In particular see Martin \textit{et al.}\cite{Martin07}, Duan \textit{et al.} \cite{LianMartinEnthalpy,LianMartinMach,duan_beekman_martin_2010,TaylorMartin07}.\\
\indent For the statistical analysis, time-domain signals of primitive flow variables are collected from the outlet plane, $x_o/\delta$, of each run. The total sample time in outer time units $t(\overline{u}_e/\delta)$ is provided in Table~\ref{tab:02}. No statistics are presented from the auxiliary boundary layer runs (M7T5* and M12T5*). The incompressible channel flow DNS database of Lee \&  Moser \cite{LeeMoser15} and Bernardini \& Pirozzoli \cite{BOP} with $Re_{\tau}$ ranging approximately from 1000 to 5200 is listed in table \ref{Incompressible DNS Database} to be used as reference IBTL cases.\\

\subsection{Averaging and notation}
\indent In this paper, both Reynolds and Favre averaged quantities are employed. As noted previously, Reynolds-averaged quantities are denoted by an overbar, $\overline{f}$, with a single prime denoting a fluctuation relative to this average, $f^\prime$. Similarly, Favre averaged quantities are denoted by tilde, $\Tilde{f}$, and fluctuations from this average are, $f''$. The variable $f$ represents a flowfield variable such as the streamwise, spanwise or wall-normal velocities which we denote as $u$, $v$ and $w$, respectively. While dataset conditions and the initial thermodynamic fluctuation analysis  will be predominantly described in terms of Reynolds averaged quantities, the majority of flowfield statistics and the following derivation of an MVT for compressible flows will be conducted in density-weighted Favre-averaged variables. The following work can be conducted fully with Reynolds or Favre averaging without impacting our conclusions.


\begin{sidewaystable}
    \caption{Boundary layer edge and wall parameters of DNS database.}
    \smallskip
    \label{Compressible DNS Database}    
    \begin{tabular}{||lcccccccccccccc||}
    \hline
        Case        & Viscosity Law & $M_e$ & $\overline{u}_e$& $\overline{T}_e$&$\overline{T}_w/\overline{T}_r$ & $-Bq$ & $Re_\theta$ & $Re_{\delta 2}$ & $Re_\tau$ & $Re^*$ & $u_{\tau}$    &$\overline{\rho}_w$ & $\overline{\rho}_e$ & line\&symbol
        \\[3pt]
                    &               &       & m/s   &     K &     & &          &          & &  & $m\cdot s^{-1}$    &   $kg\cdot m^{-3}$ &$kg \cdot m^{-3}$ &\\         
        \hline
        M3T5     &  Power       & 3.0   & 882.5 &   220     &1.0     &  0.0 & 3480      & 1760            & 530       & 1650   &        44.5    &   0.0366     &       0.0917  &  \textcolor[rgb]{0.4660 0.6740 0.1880} {$\sampleline{}\quad \bullet$} \\ 
        M5T5     &  Power       & 4.9   & 1472  &   225     &1.0     &  0.0 & 7450      & 1980            & 470       & 3870   &        81.5    &   0.0182     &       0.0968  &  \textcolor[rgb] {1 0 1} {$\sampleline{} \quad  \bullet$}\\
        M7T5-L   &  Power       & 6.9   & 2069  &   224     &1.0     &  0.0 & 15600      & 2830            & 550       & 9450   &       115.9    &   0.00969    &   0.0926 &  \textcolor[rgb]{0 0 1} {$\sampleline{}\quad \bullet$}\\
        M12T5-L  &  Power       & 11.7  & 3612  &   236     &1.0     &  0.0 & 46800      & 3880            & 550       & 33990  &       213.1    &   0.0038     &   0.1015 &  \textcolor[rgb]{0 1 1} {$\sampleline{}\quad \bullet$}\\
        M5T3     &  Power       & 4.9   & 1477  &   222     &0.5     &  0.05 & 4650      & 2054            & 610       & 2350   &       68.0    &   0.0324     & 0.0950 &  \textcolor[rgb]{1 0 0} {$\sampleline{}\quad \blacklozenge$} \\
        M5T1     &  Power       & 5.0   & 1498  &   223     &0.2     & 0.17 & 1620      & 1650            & 830       & 800    &       48.4    &   0.0973     &   0.0951 &  \textcolor[rgb]{0 1 0} {$\sampleline{}\quad \bigstar$} \\
        M10T3    &  Keyes(N2)   & 9.1   & 1410  &   58.6    &0.5     & 0.11 & 7565      & 1745            & 491       & 4827   &       63.0    &   0.0079     &    0.0403 &  \textcolor[rgb]{0.4940 0.1840 0.5560} {$\sampleline{}\quad \blacklozenge$} \\  
        \hline
    \end{tabular}
    \bigskip
    \caption{\raggedright Computational domain size and grid resolution for the DNS data. Datasets with case name ending in `-L' are long-box runs with the domain length extended utilizing starred (*) auxiliary simulations at their inlet.  All other cases use the rescaling method of Xu \& Martin\cite{XuMartin04} for inflow assignment. The $\sim$ indicates the cumulative sampling distance once auxiliary simulation distances are taken into account.}
    \smallskip
    \begin{tabular}{||lccccccccccccc||}
      \hline
        Case  & $N_x$ & $N_y$ & $N_z$    & $\delta$ & $L_x/\delta$ & $L_y/\delta$ & $L_z/\delta$ & $\Delta x^+$ & $\Delta y^+$ & $z_2^+$ & $\chi$ & $t \overline{u}_e/\delta$ & $x_o/\delta$\\
        
                &&&& $mm$     &       &      &       &      &       &       &         &        &   \\[3pt]
                \hline
        M3T5    & 1820 &880 & 110      & 9.3      & 28.7  & 5.7  & 7.5   & 8.3  & 3.4   & 0.32  & 1.063   & 143   &  26.8 \\
        M5T5    & 1820 &880 & 110      & 16.6     & 27.2  & 5.4  & 8.1   & 7.1  & 2.9   & 0.26  & 1.061   & 134   &  26.1 \\
        M7T5*   & 1780 & 1160 & 110     & 39.8        & 21.1  & 5.6  & 6.12  & 6.6  & 2.7   & 0.26  & 1.061   &  134     & 20.2  \\
        M7T5-L  & 1580 & 1080 & 116     & 39.8     & 21.0  & 5.6  & 7.6   & 7.4  & 2.9   & 0.28  & 1.061   & 110   &  40.2 $\sim$ \\
        M12T5*  & 1640 & 1300 & 110     & 125.8        & 20.4  & 5.4  & 5.8   & 6.8  & 2.4   & 0.28  & 1.060   & 110      &  19.0 \\
        M12T5-L & 1640 & 1240 & 116     & 125.8    & 20.4  & 5.4  & 7.2   & 6.8  & 2.4   & 0.29  & 1.060   & 142   &  38.7$\sim$ \\
        M5T3    & 2032 & 1080 & 106     & 9.0      & 25.4  & 5.1  & 8.4   & 7.6  & 2.9   & 0.32  & 1.069   & 199   &  24.0 \\
        M5T1    & 2080 & 1648 & 110     & 2.5      & 19.5  & 5.8  & 7.3   & 7.7  & 2.9   & 0.29  & 1.069   & 190   &  18.5 \\
        M10T3   & 1920 & 1680 & 112     & 17.8     & 30.3  & 10.2 & 10.7  & 7.8  & 2.9   & 0.31  & 1.065   &  93     &   29.0\\
        \hline
    \end{tabular}
    \label{tab:02}
\end{sidewaystable}

\begin{table} 
\centering
\caption{Incompressible Channel Flow DNS Database}\label{Incompressible DNS Database}
\begin{tabular}{|| c c c c ||}
Case & $Re_\tau$& line\&symbol & Reference  \\
\midrule
 LM5200    & 5186  &  \textcolor[rgb]{0 0 0} {$\sampleline{}\quad \blacksquare$} & Lee \& Moser (2015)\cite{LeeMoser15} \\
 LM2000    & 1994  &  \textcolor[rgb]{0 0 0} {$\sampleline{dashed}\quad \blacklozenge$} & Lee \& Moser (2015)\cite{LeeMoser15}\\
 LM1000    & 1000  &  \textcolor[rgb]{0 0 0} {$\sampleline{dotted}\quad \bigstar$} & Lee \& Moser (2015)\cite{LeeMoser15}\\
 BOP4100   & 4079  &  \textcolor[rgb]{0 0 0} {$\sampleline{dash dot}\quad \bullet$} & Bernardini and Pirozzoli (2014)\cite{BOP}\\
\bottomrule
\end{tabular}
\end{table}

\section{Density and Viscosity Fluctuations in Compressible Boundary Layers}
\subsection{Breakdown of the Classical Momentum Balance}
\indent To probe influences of compressibility on thermodynamic property fluctuations on compressible turbulent boundary layers, we begin by examining the Reynolds-averaged, thin shear layer stream-wise momentum equation. While it is customary to use Favre-averaged equations in compressible flow, the use of the Reynolds-averaged equation provides insights into the influence of fluid property variations in compressible turbulence that are rarely remarked upon. We begin  by employing the classical assumptions that the stream-wise derivatives are negligible, the fluctuations of thermodynamic variables are small, the pressure gradient across the boundary layer is small and the magnitude of convection terms are negligible near the wall. Under these conditions, the momentum equation reduces to,

\begin{equation}
    \label{equation:ThinShearEq}
        1 \simeq \frac{\overline{\mu}}{\tau_w}\frac{\partial \overline{u}}{\partial z} - \frac{\overline{\rho}\overline{u'w'}}{\tau_w}=\tau_{V-R}^++\tau_{T-R}^+
\end{equation}

\begin{figure*}
\centering
\begin{subfigure}{.5\textwidth}
  \centering
  \includegraphics[width=\linewidth]{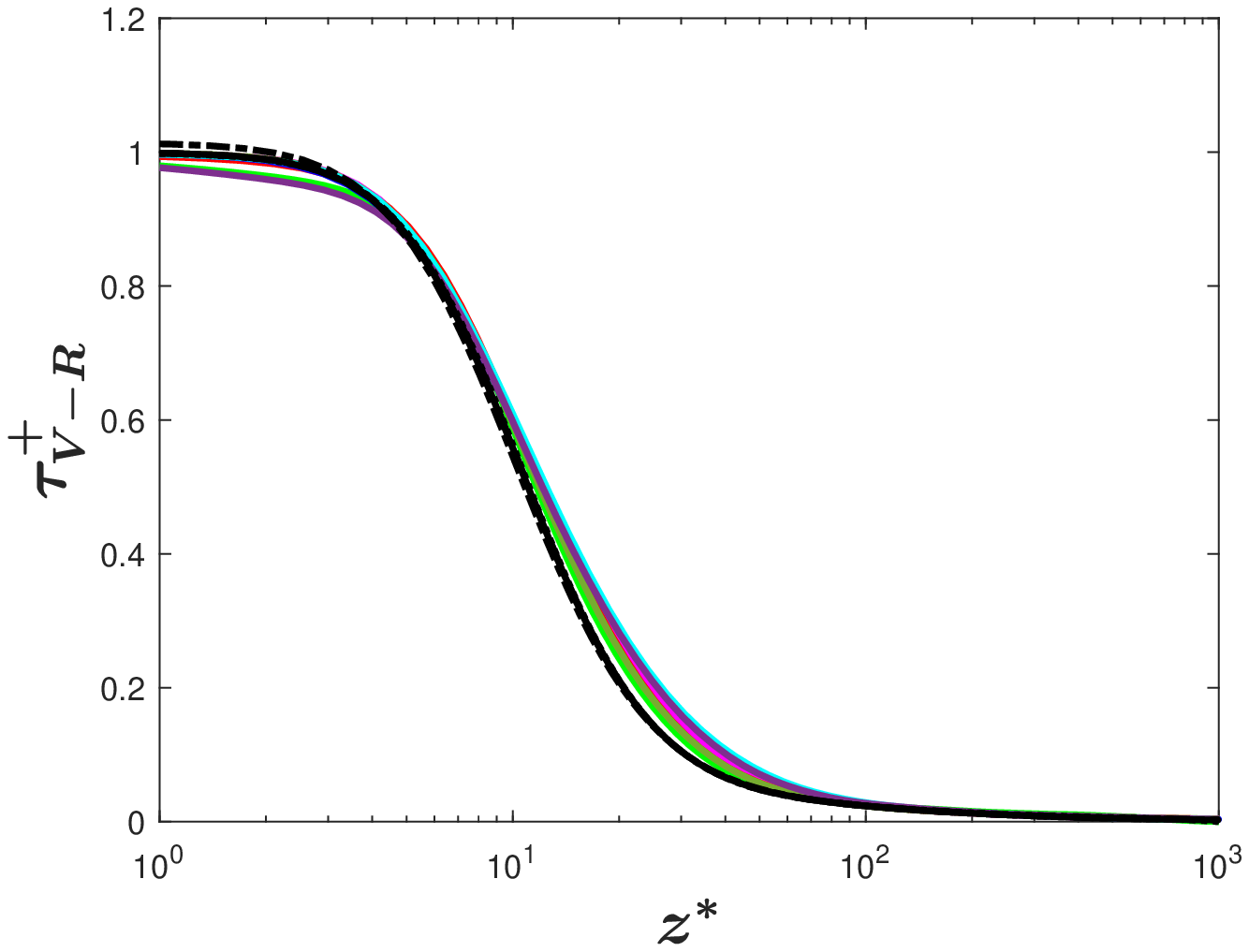}
  \caption{}
  \label{fig:uncorrected_VS}
\end{subfigure}%
\begin{subfigure}{.5\textwidth}
  \centering
  \includegraphics[width=\linewidth]{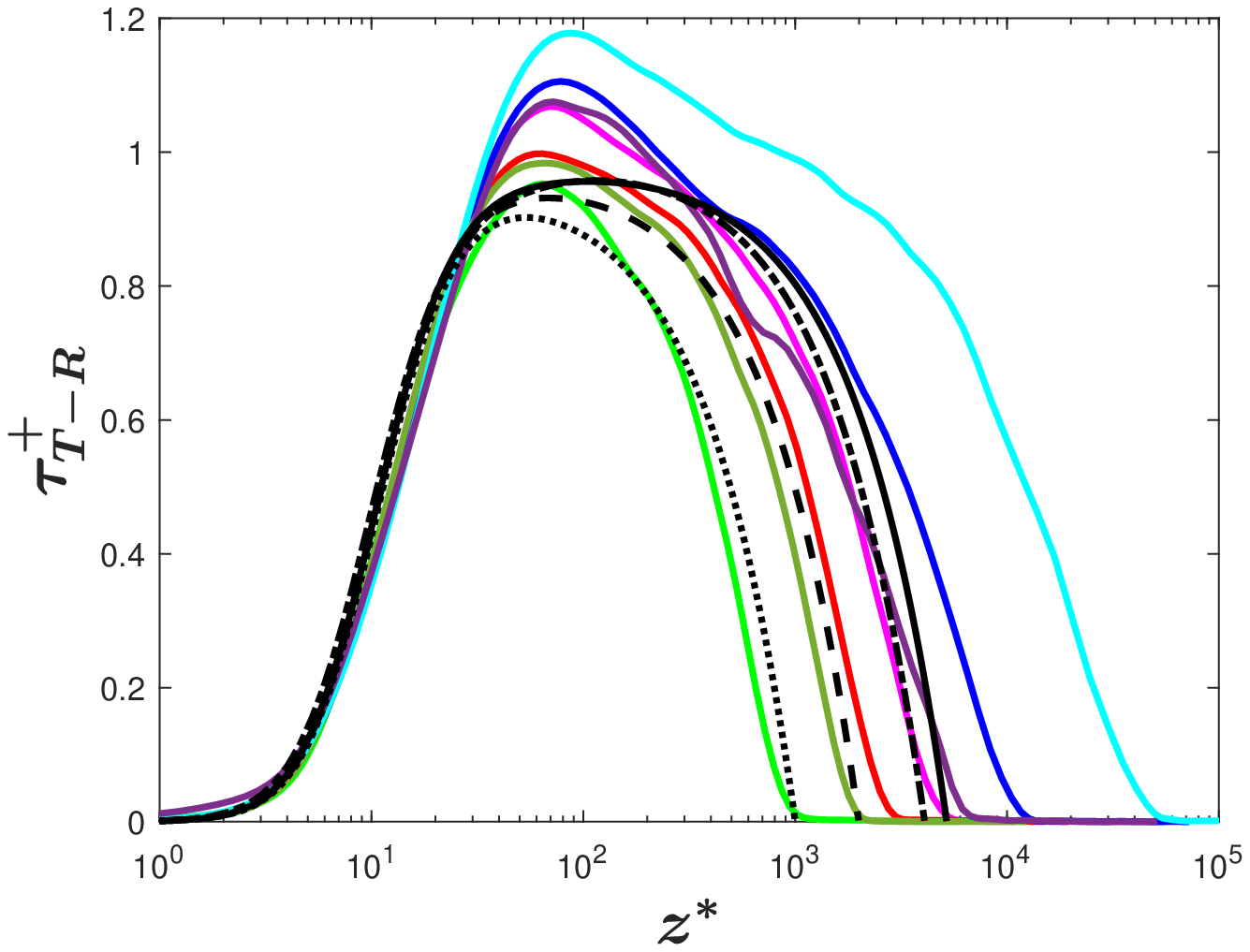}
  \caption{}
  \label{fig:Uncorrected_RS}
\end{subfigure}
\begin{subfigure}{.5\textwidth}
  \centering
  \includegraphics[width=\linewidth]{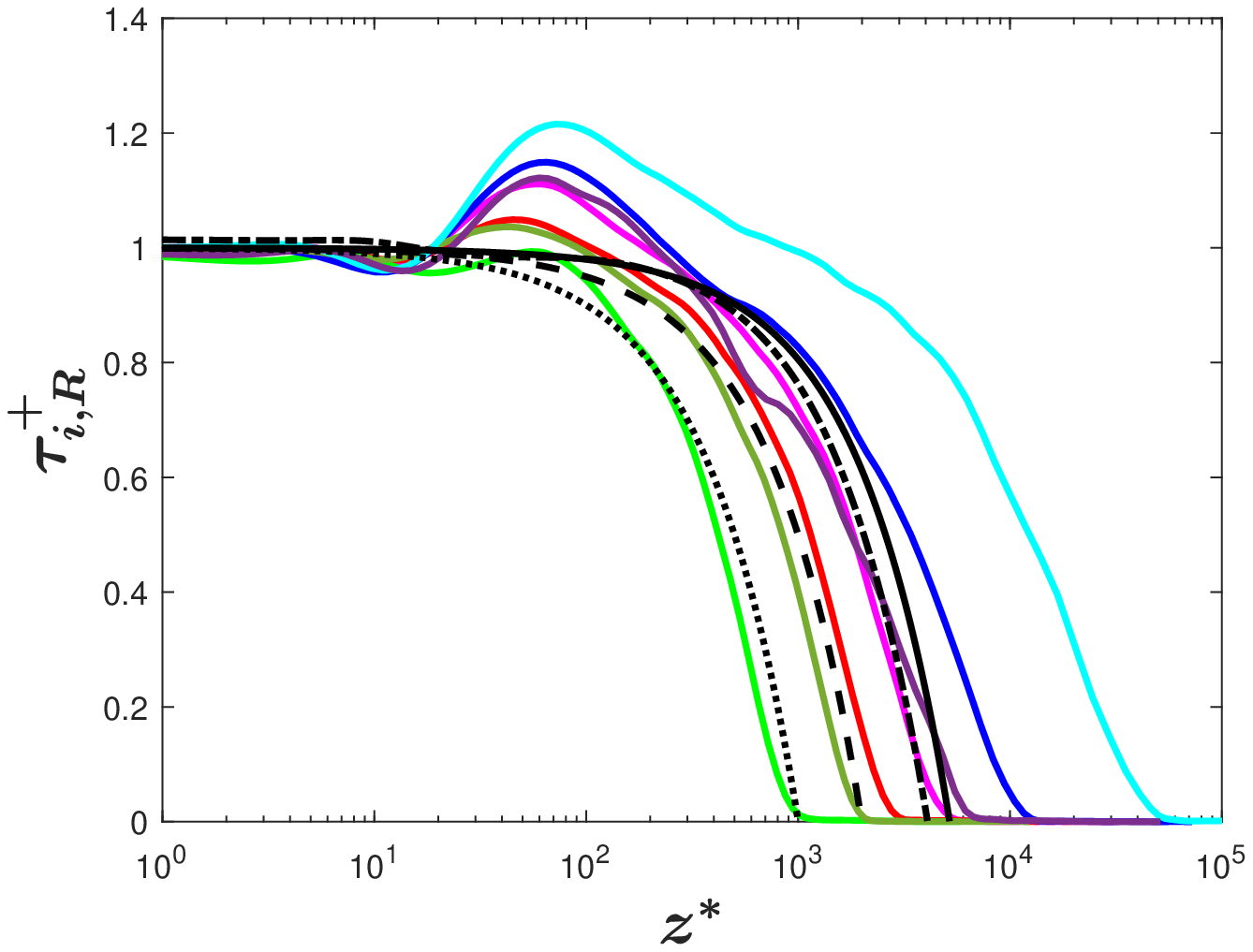}
  \caption{}
  \label{fig:uncorrected_TS}
\end{subfigure}
\caption{Non-dimensionalized (A) viscous stress, $\tau_{V-R}^+$, (B) turbulent stress, $\tau_{T-R}^+$, (C) total stress, $\tau_{i,R}^+$, are plotted against the semi-local wall normal coordinate, $z^*$. Corresponding stresses from ITBL flow references are included for comparison. References for lines colors and styles as well as references for the database are included in Table \ref{Compressible DNS Database} and \ref{Incompressible DNS Database}}
\label{fig:UncorrectedStresses}
\end{figure*}
\begin{figure*}
\centering
\begin{subfigure}{.5\textwidth}
  \centering
  \includegraphics[width=\linewidth]{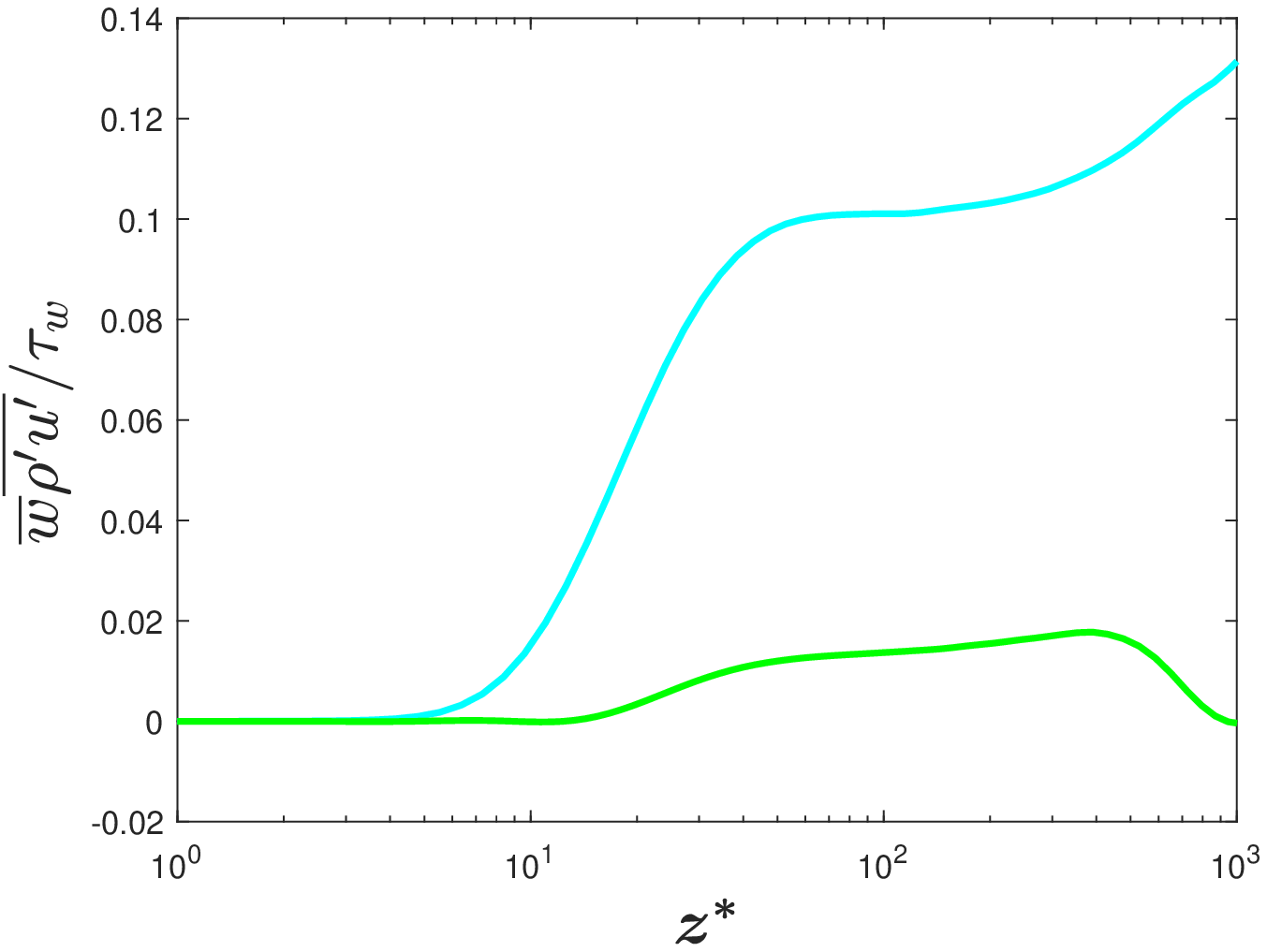}
  \caption{}
  \label{fig:waverpup}
\end{subfigure}%
\begin{subfigure}{.5\textwidth}
  \centering
  \includegraphics[width=\linewidth]{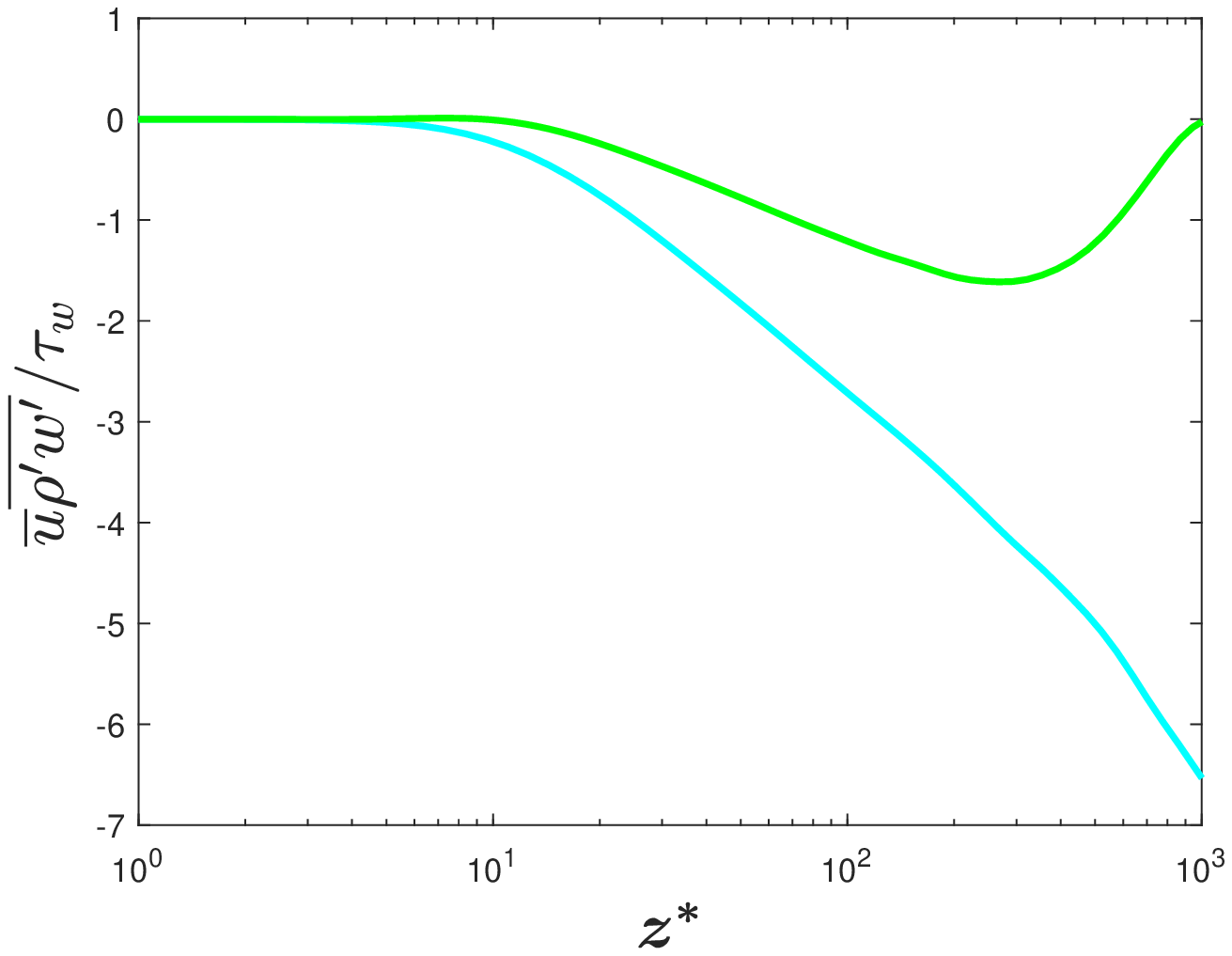}
  \caption{}
  \label{fig:uaverpwp}
\end{subfigure}
\begin{subfigure}{.5\textwidth}
  \centering
  \includegraphics[width=\linewidth]{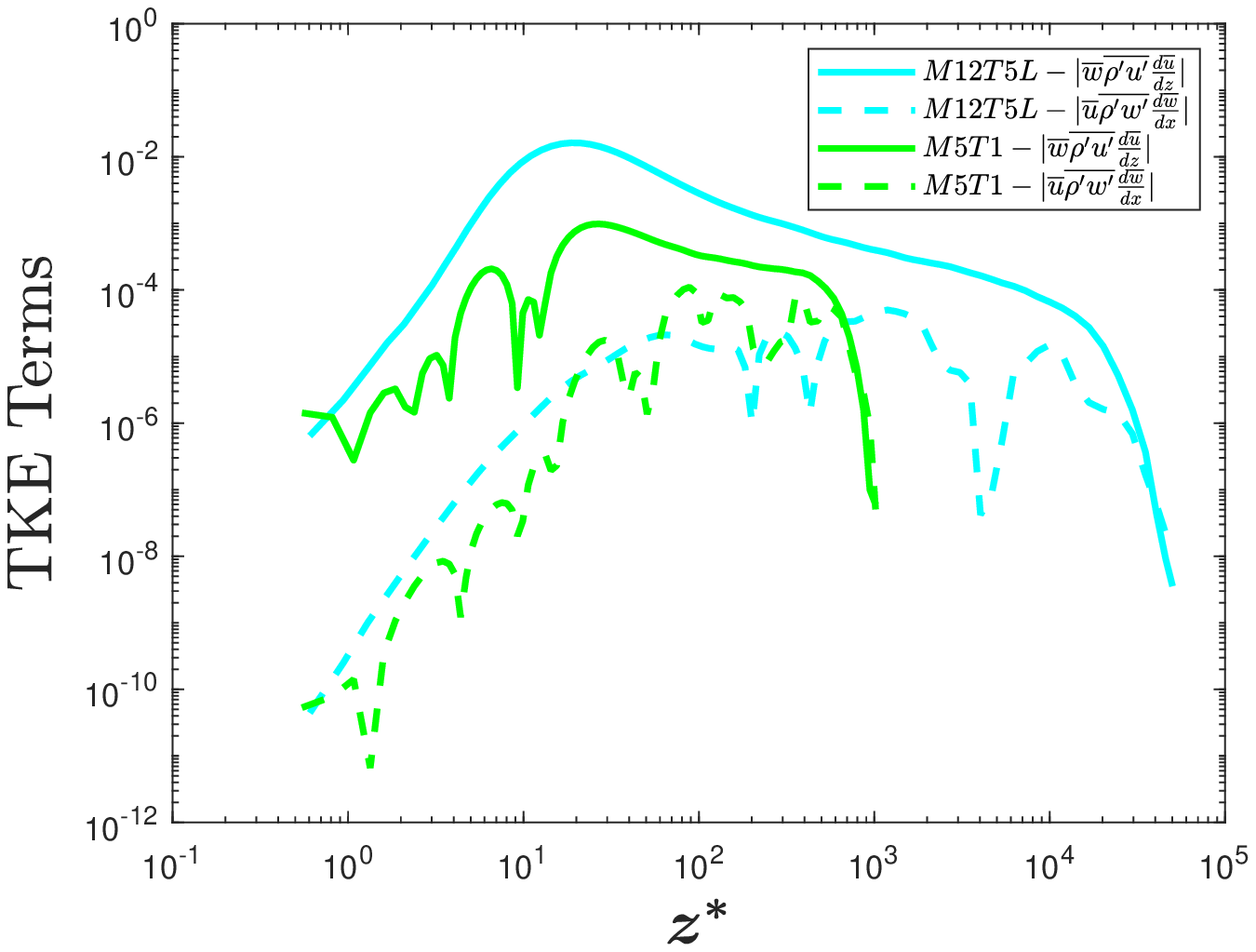}
  \caption{}
  \label{fig:selectTKE}
\end{subfigure}%
\caption{\textcolor{blue}{Comparison of 
 (A) $\overline{w}\overline{\rho'u'}/\tau_w$ and (B) $\overline{u}\overline{\rho'w'}/\tau_w$ for M5T1 and M12T5. (C) Comparison of the magnitudes of the TKE terms corresponding to $\overline{w}\overline{\rho'u'}$ and $\overline{u}\overline{\rho'w'}$. All TKE terms are normalized by $\overline{\rho}u_\tau^{*3}/z_\tau^*$ where $u^*_\tau = u_\tau\sqrt{\rho_w/\overline{\rho}}$ and $z_\tau^* = \overline{\mu}/\overline{\rho}u_\tau\sqrt{\rho_w/\overline{\rho}}$. References for line colors and symbols are included in Table \ref{Compressible DNS Database}.}}
\label{fig:selectTKE_convectionlike}
\end{figure*}

\noindent where the superscript "+" denotes the non-dimensionalization by  wall quantities, and $\tau_{V-R}^+$ and $\tau_{T-R}^+$ are the non-dimensional Reynolds-averaged viscous and turbulent stresses, respectively. The classical momentum balance described by Eq.\ref{equation:ThinShearEq} is expected to hold near the wall for zero pressure gradient boundary layers if Morkovin's hypothesis is valid; that is, the density and viscosity fluctuations are negligible and the consideration of only mean density and mean viscosity is sufficient to restore the classical near wall momentum balance observed in ITBLs. \\
\indent The stress balance and the magnitude of its classic viscous and turbulent contributions is considered in Fig.~\ref{fig:UncorrectedStresses} \, where $\tau_{i,R}^+=\tau_{V-R}^++\tau_{T-R}^+$ denotes the non-dimensionalized total stress. The subscript 'R' is used to denote the use of Reynolds-averaging. As expected for the incompressible cases, the classical momentum balance holds in the near wall region, with the incompressible total stress in Fig.~\ref{fig:uncorrected_TS} is seen to be very close to one. In contrast, the balance is not observed for many CTBL datasets, which is most clearly demonstrated by the turbulent shear stress achieving magnitudes of up to 1.2 times the wall shear stress (see Fig.~\ref{fig:Uncorrected_RS}). This overshoot in the wall shear stress has been observed in a number of previous compressible wall-bounded studies ~\cite{LianMartinEnthalpy,LianMartinMach,ZhangLian2018,lagha-kim-zhong-2011}, suggesting a violation of the underlying assumptions inherent in the classical near-wall momentum balance described by Eq.\ref{equation:ThinShearEq}.  \\
\indent Extending the TSL Reynolds averaged momentum equation to include terms involving density and viscosity fluctuations, we can arrive at  Eq. \ref{equation:AlltermsdensityTSL}. 
\begin{equation}
\label{equation:AlltermsdensityTSL}
    1 = \tau_{V-R}^++\tau_{T-R}^+ + \left(\frac{-\overline{u}\overline{\rho'w'}-\overline{w}\overline{\rho'u'}-\overline{\rho'u'w'}}{\tau_w}\right) + \left(\frac{\overline{\mu'\frac{\partial u'}{\partial z}}}{\tau_w}\right)  
\end{equation}
\noindent which involves three additional non-dimentionalized density fluctuation terms and one more viscosity fluctuation term. \textcolor{blue}{In particular, $\overline{w}\overline{\rho'u'}$ and $\overline{u}\overline{\rho'w'}$ are of special interest. The magnitude of the two terms is compared in Fig.~\ref{fig:selectTKE_convectionlike} relative to the wall shear stress for the  M5T1 and M12T5 cases. These cases have been selected as they are representative of the range of these terms observed in this dataset. The $\overline{u}\overline{\rho'w'}$ term is seen to be much larger than $\overline{w}\overline{\rho'u'}$ and can even be as large as 3 times the wall shear stress at $z^* = 100$. However, we also observe in Fig.~\ref{fig:waverpup} that $ \overline{w}\overline{\rho'u'}$ accounts for 2\% to 10\% of wall shear stress for M5T1 and M12T5 case, respectively, and cannot be neglected as has traditionally been assumed (see Spina et al.\cite{SpinaSmitsRobinson}). }


\textcolor{blue}{The large magnitude of $\overline{u}\overline{\rho'w'}$ shown in Fig.~\ref{fig:uaverpwp} is challenging, however, because it is so large in some parts of the layer that its inclusion as a term in the near-wall stress balance of Eq.~\ref{equation:AlltermsdensityTSL} would cause the right hand side of Eq. \ref{equation:AlltermsdensityTSL} to increase significantly above the wall shear stress outside of the near-wall region; significantly different behavior and trends than observed in Fig \ref{fig:UncorrectedStresses}.} 
\textcolor{blue}{However, Fig.~\ref{fig:selectTKE} reveals that $\overline{u}\overline{\rho'w'}d\overline{w}/dx$, a corresponding TKE term to $\overline{u}\overline{\rho'w'}$, is orders of magnitude smaller than $\overline{w}\overline{\rho'u'}d\overline{u}/dz$, a corresponding TKE term to $\overline{w}\overline{\rho'u'}$, suggesting that despite the large magnitude of $\overline{u}\overline{\rho'w'}$ in Eq.\ref{equation:AlltermsdensityTSL}, its role in the transport of turbulent kinetic energy and turbulent momentum is limited. The above TKE transport argument closely follows that of Spina et al.\cite{SpinaSmitsRobinson}. They interpreted $\overline{u}\overline{\rho'w'}$ as the "mean rate of transfer of turbulent mass flux across the same plane" and asserted that the term does not involve the transport of turbulent momentum. In addition, Spina et al.\cite{SpinaSmitsRobinson} suggested that terms important to turbulent kinetic energy are two orders of magnitude greater than the TKE term corresponding to $\overline{u}\overline{\rho'w'}$ and thus, neglected $\overline{u}\overline{\rho'w'}$ based on that reasoning. Overall, Spina et al.\cite{SpinaSmitsRobinson} neglected $\overline{w}\overline{\rho'u'}$ by arguing that its magnitude within the momentum equation was small in many cases. We have demonstrated in Fig.~\ref{fig:waverpup} that $\overline{w}\overline{\rho'u'}$ is non-negligible for our dataset that includes high Mach and Re* cases and thus choose to retain the term in Eq.~\ref{equation:DensityCorrectedTSLwithZeta}. However, we follow the similar argument as Spina et al.\cite{SpinaSmitsRobinson} when neglecting $\overline{u}\overline{\rho'w'}$ by demonstrating the small magnitude of the corresponding TKE transport term.}

\textcolor{blue}{Another approach to examine the treatment of $\overline{u}\overline{\rho^\prime w^\prime}$ explores the near-wall momentum equation prior to neglecting one of the convective terms, as seen in Eq.  \ref{equation:AlltermsdensityTSLderivation} (essentially a precursor to Eq. ~\ref{equation:AlltermsdensityTSL}.) }

\textcolor{blue}{
\begin{equation}
\label{equation:AlltermsdensityTSLderivation}
     \underbrace{\frac{d}{dz^*}\left[ \frac{\overline{\rho}\overline{u}\overline{w} + \overline{u}\overline{\rho^\prime w^\prime}}{\tau_w}\right]}_{\simeq 0}= \frac{d}{dz^*}\left[ \left(\overline{\mu}\frac{d\overline{u}}{dz}-\overline{\rho}\overline{u^\prime w^\prime} + \left(-\overline{w}\overline{\rho'u'}-\overline{\rho'u'w'}\right) + \left(\overline{\mu'\frac{\partial u'}{\partial z}}\right)\right)/\tau_w \right] 
\end{equation}}

\begin{figure*}
\centering
\begin{subfigure}{.5\textwidth}
  \centering
  \includegraphics[width=\linewidth]{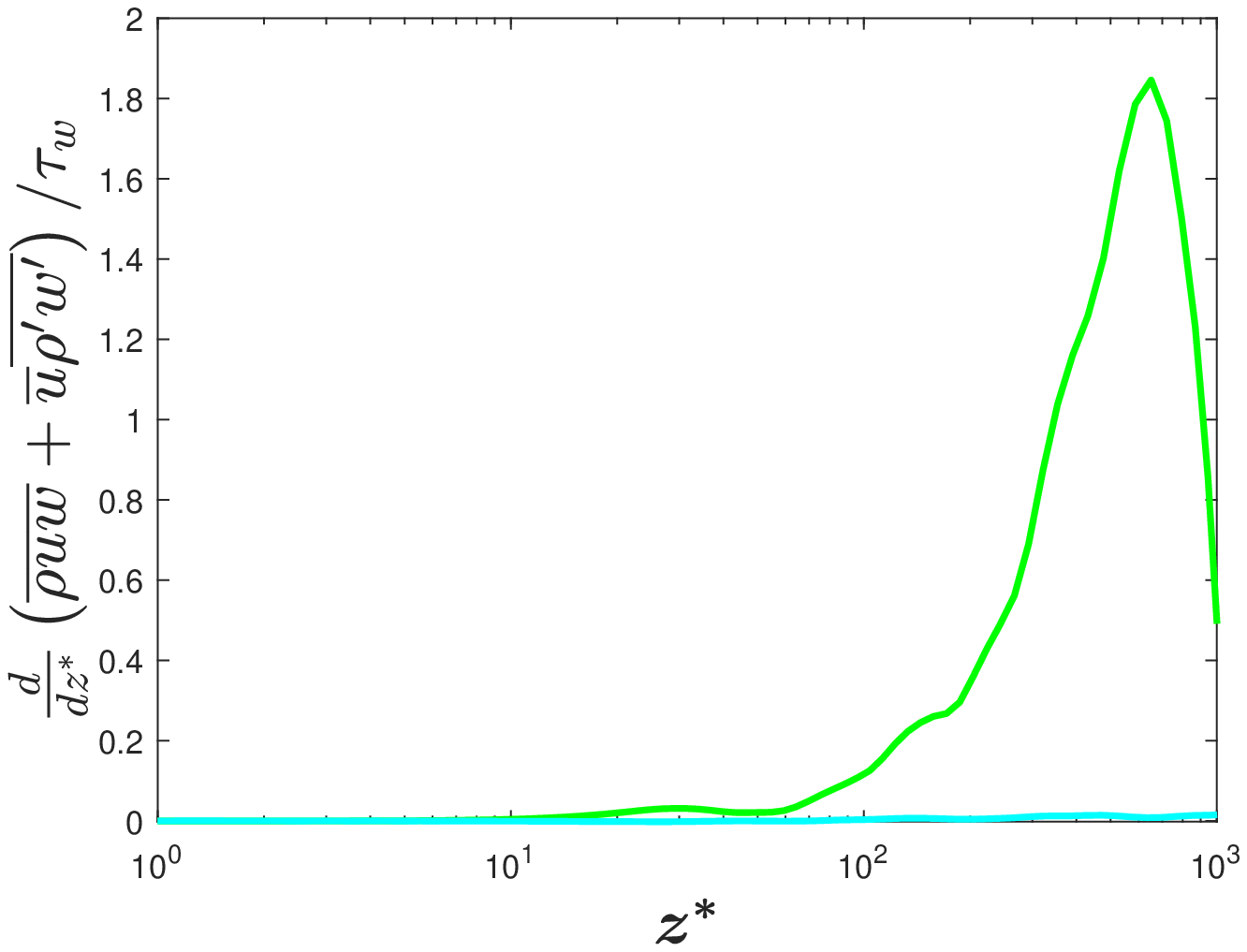}
  \caption{}
  \label{fig:convection}
\end{subfigure}%
\begin{subfigure}{.5\textwidth}
  \centering
  \includegraphics[width=\linewidth]{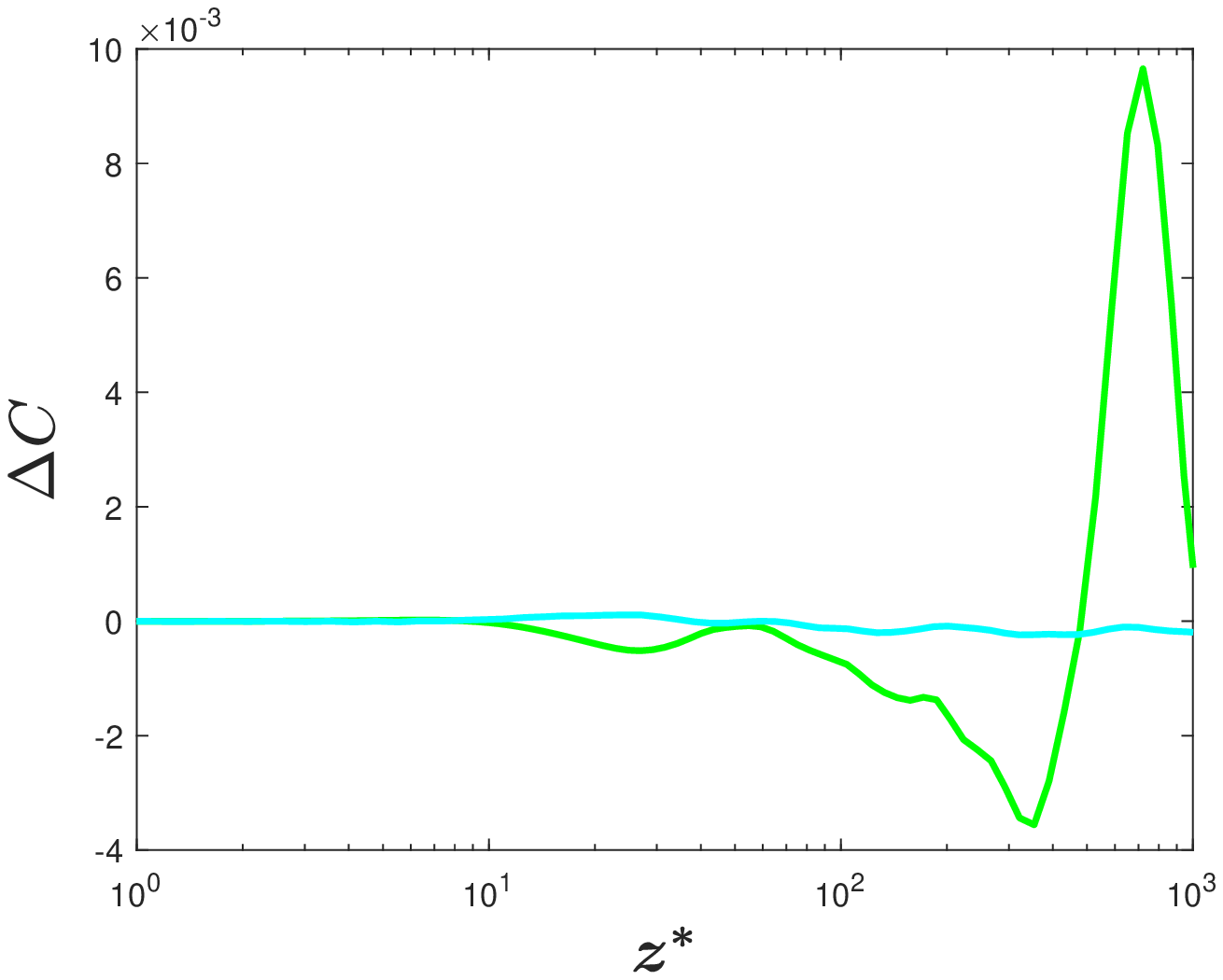}
  \caption{}
  \label{fig:convectiondiff}
\end{subfigure}
\caption{\textcolor{blue}{(A) Left hand side of Eq.\ref{equation:AlltermsdensityTSLderivation}, $\frac{d}{dz^*}\left[ \overline{\rho}\overline{u}\overline{w} + \overline{u}\overline{\rho^\prime w^\prime}\right]/\tau_w$. (B) Difference between $\overline{\rho}\tilde{u}\tilde{w}$ and $\overline{\rho}\ \overline{u}\ \overline{w} + \overline{u}\overline{\rho^\prime w^\prime}$ normalized by $\tau_w$, where $\Delta C = \overline{\rho}\tilde{u}\tilde{w}/\tau_w- \left(\overline{\rho}\ \overline{u}\ \overline{w} + \overline{u}\overline{\rho^\prime w^\prime}\right)/\tau_w$. References for line colors are included in Table \ref{Compressible DNS Database}. }}
\label{fig:convection&diff}
\end{figure*}

\textcolor{blue}{The left hand side of Eq.~\ref{equation:AlltermsdensityTSLderivation} is plotted  in Fig.~\ref{fig:convection}. It can be observed that, for both M5T1 and M12T5 cases, which are representative of the range of this term in this dataset, 
the left hand side of Eq.~\ref{equation:AlltermsdensityTSLderivation} remains close to zero through much of the layer. Moreover, it can be shown that the difference between $\overline{\rho}\ \overline{u}\ \overline{w} + \overline{u}\overline{\rho^\prime w^\prime}$  and $\overline{\rho}\tilde{u}\tilde{w}$ are shown to be minimal (Fig.~\ref{fig:convectiondiff}), suggesting that $\overline{\rho}\ \overline{u}\ \overline{w} + \overline{u}\overline{\rho^\prime w^\prime}$ is similar to the Favre averaged convection term, which is often neglected in the TSL form of the Favre momentum equation. Therefore, neglecting the $\overline{u}\overline{\rho^\prime w^\prime}$ in the Reynolds averaged stress balance in Eq.~\ref{equation:AlltermsdensityTSL} makes the two averaging approaches consistent in terms of what is neglected. 
Overall, based on the above analysis of the corresponding TKE terms and convective term analysis in Fig~\ref{fig:selectTKE} and Fig.~\ref{fig:convection&diff}, respectively we neglect $\overline{u}\overline{\rho^\prime w^\prime}$ when exploring the near wall momentum balance utilizing Reynolds averaging. This approach is consistent with the arguments of Spina et al\cite{SpinaSmitsRobinson}. }

\begin{figure}
\centering
\begin{subfigure}{.47\textwidth}
  \centering
  \includegraphics[width=\linewidth]{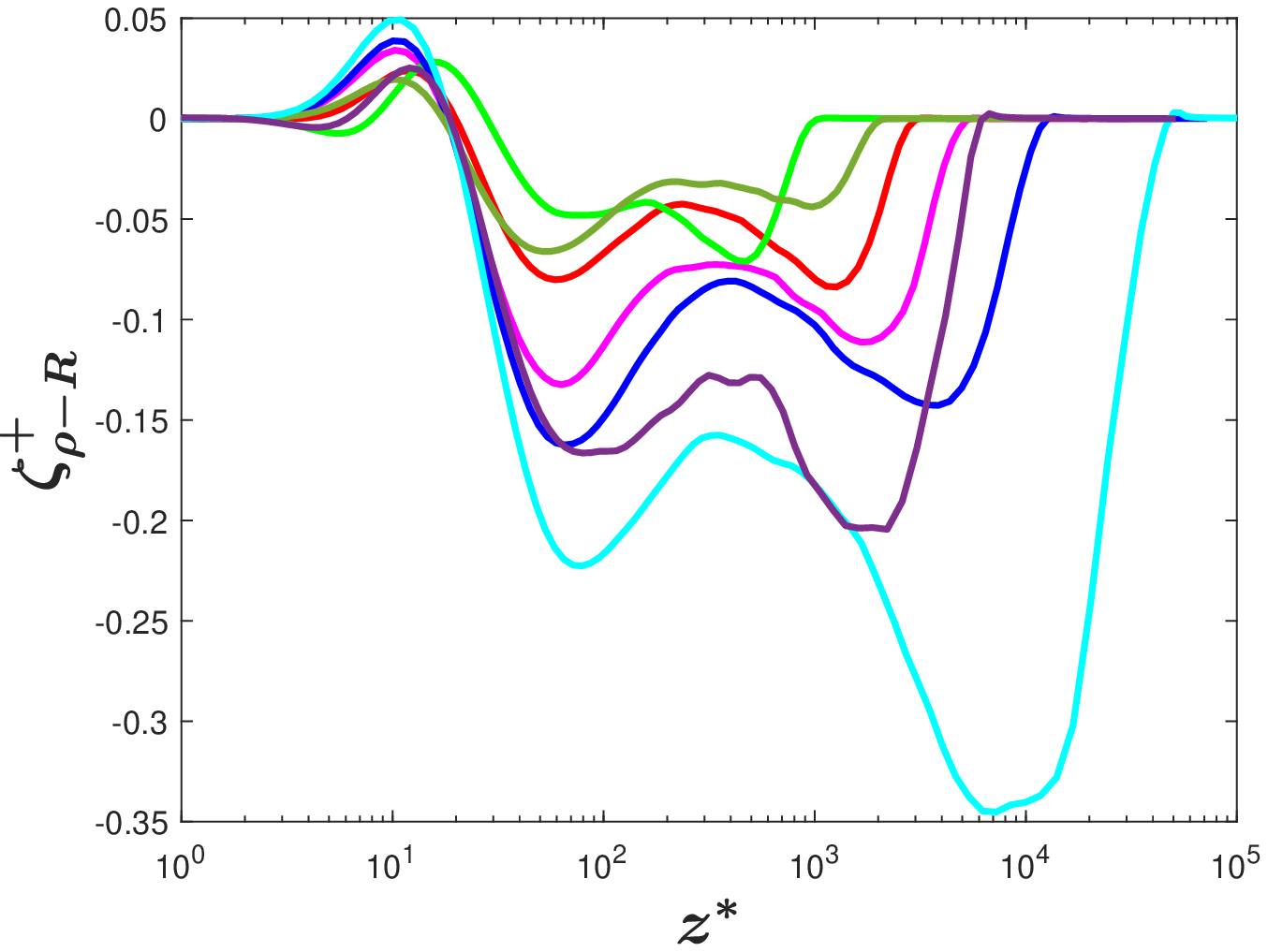}
  \caption{}
  \label{fig:zeta_rho}
\end{subfigure}
\begin{subfigure}{.47\textwidth}
  \centering
  \includegraphics[width=\linewidth]{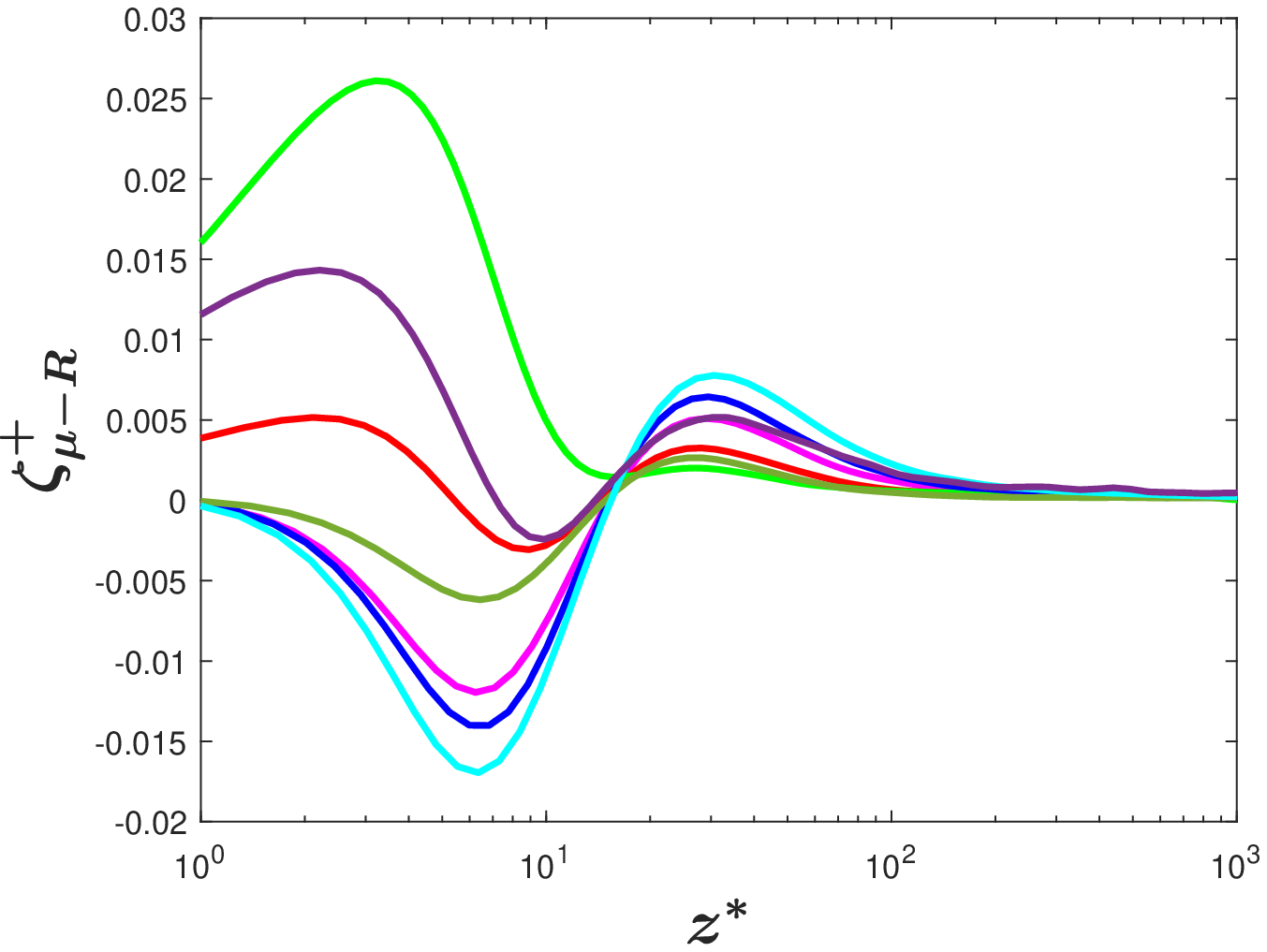}
  \caption{}
  \label{fig:zeta_vis}
\end{subfigure}
\begin{subfigure}{.47\textwidth}
  \centering
  \includegraphics[width=\linewidth]{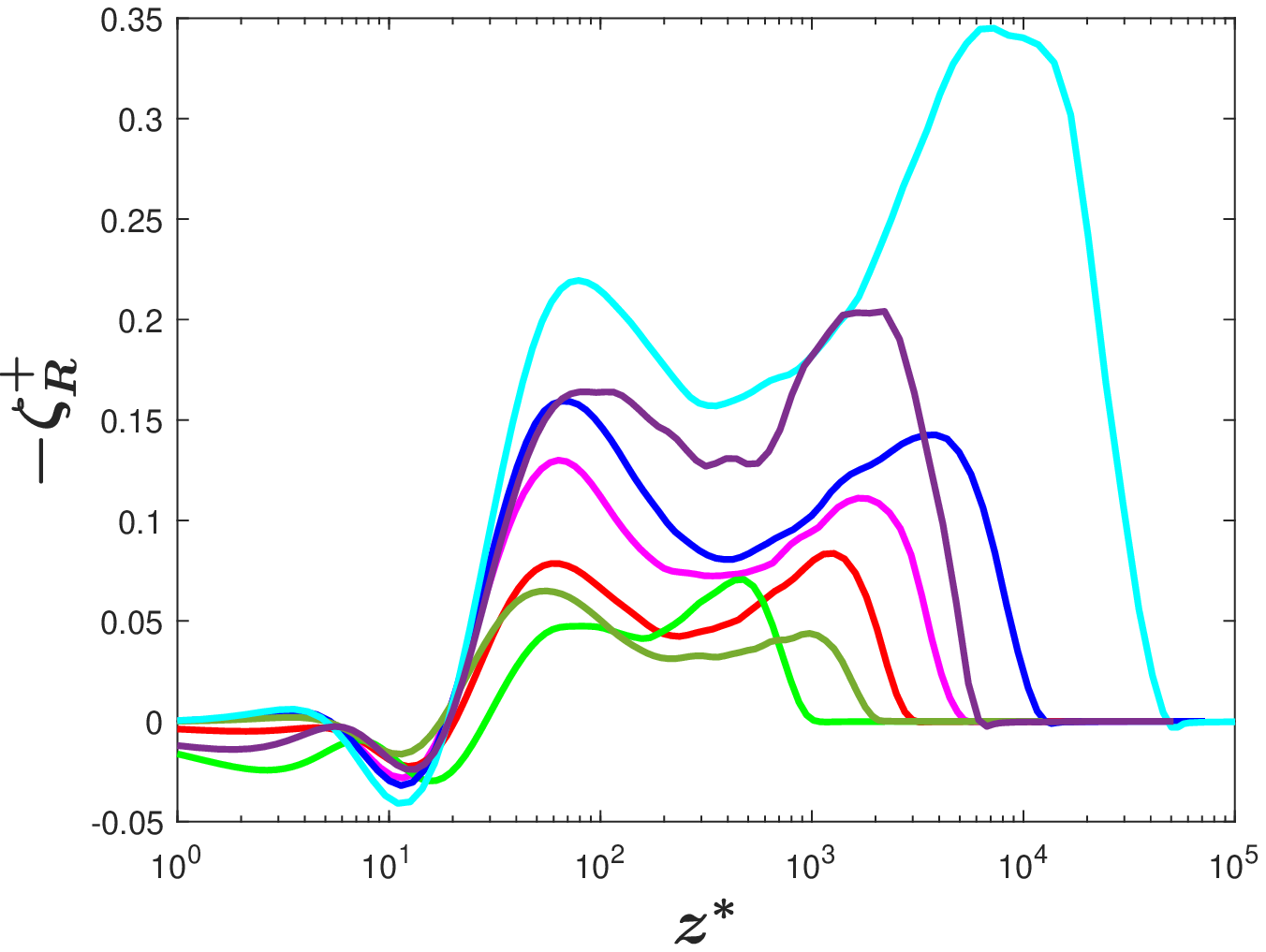}
  \caption{}
  \label{fig:zetaTot}
\end{subfigure}
\begin{subfigure}{.47\textwidth}
  \centering
  \includegraphics[width=\linewidth]{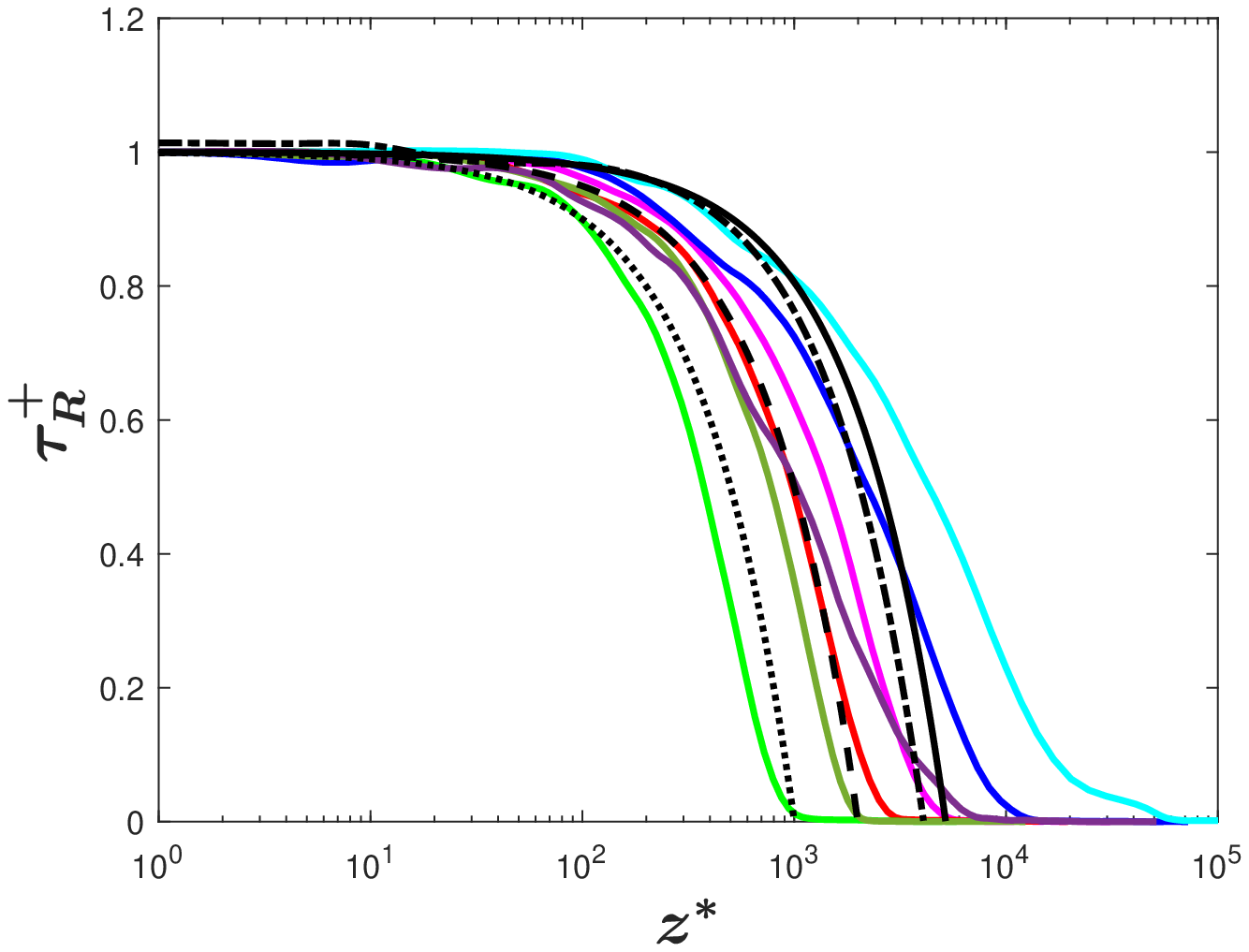}
  \caption{}
  \label{fig:tau_TS_R}
\end{subfigure}
  \caption{(A) Sum of density fluctuating term, $\zeta_{\rho-R}^+$, (B) Sum of viscosity fluctuating terms, $\zeta_{\mu-R}^+$, (C) Sum of the density and viscosity fluctuation terms, $\zeta^+_{R}$, versus semi-local wall normal coordinate, $z^*$. (D) Density and viscosity fluctuation corrected Reynolds-averaged total stress, $\tau^+_{R}$. References for line colors and styles as well as references for the database are included in table \ref{Compressible DNS Database}.}
  \label{fig:zeta}
\end{figure}

\indent The resulting equation is shown in Eq.\ref{equation:DensityCorrectedTSLwithZeta} which now involves two non-dimentionalized density fluctuation terms and one viscosity fluctuation term that we group as ${\zeta}_{\rho-R}^+$ and $\zeta_{\mu-R}^+$, respectively.  
\begin{equation}
\label{equation:DensityCorrectedTSLwithZeta}
    1 = \tau_{V-R}^++\tau_{T-R}^+ + \left(\underbrace{\frac{-\overline{w}\overline{\rho'u'}-\overline{\rho'u'w'}}{\tau_w}}_{\zeta^+_{\rho-R}}\right) + \left(\underbrace{\frac{\overline{\mu'\frac{\partial u'}{\partial z}}}{\tau_w}}_{\zeta^+_{\mu-R}}\right)  
\end{equation}
Terms included in $\zeta^+_{R} = {\zeta}_{\rho-R}^+ + \zeta_{\mu-R}^+$ represent the influence of fluctuating thermodynamic properties on the total streamwise stress balance in the Reynolds-averaged momentum equation. These terms are explored in Fig.~\ref{fig:zeta}. Interestingly, the magnitude of $\zeta^+_{R}$ exceeds 4\% for all cases considered in this study, including those at Mach 3, suggesting that thermodynamic property fluctuations should be considered in the analysis of the near-wall momentum balance of most CTBLs. Noticeably, the magnitude of the first peak in Fig. \ref{fig:zeta_rho}, which is located at the lower edge of the log layer, indicates that $\zeta_{R}^+$ ranges from 4\% to 22\% of the wall shear stress for this dataset. The magnitude of $\zeta^+_{\mu-R}$ in Fig. \ref{fig:zeta_vis} is a non-negligible proportion of this, with $\zeta^+_{\mu-R}$ ranging from -1.5\% to 2.5\% in the viscous layer. Note that the magnitude of the density and viscosity fluctuation effect will be different depending on the averaging method used and, as we will see in  a subsequent section, the viscosity fluctuation proportion of the total stress is larger under Favre averaging. Not surprisingly, when the total stress with the thermodynamic property fluctuations, $\tau^+_R = \tau^+_{V-R}+\tau^+_{V-R}+\zeta^+_R$, is plotted as shown in Fig.~\ref{fig:tau_TS_R}, the near-wall momentum balance expected from the incompressible theory is restored, supporting an inclusion of $\overline{w}\overline{\rho'u'}$ while neglecting $\overline{u}\overline{\rho'w'}$ in consideration of $\zeta^+_{\rho}$.\\
\indent We thus conclude that density and viscosity fluctuations must be considered for an accurate statement of momentum balance. This influence is reduced for hypersonic datasets at low Reynolds numbers or with significant heat transfer. At low Reynolds numbers the viscous stress is a greater contribution of the total stress in the near-wall region, leading to peak turbulent stresses, $\tau_{T-R}^+$, that are less likely to exceed the wall shear stress, masking the fluctuating density and viscosity influences. We shall see below that employing Favre averaging effectively accounts for density fluctuations but not those of viscosity. Another consequence is that $\overline{\rho}\overline{u'w'}$ and $\overline{\mu}{\partial \overline{u}}/{\partial z}$ are not equal to $\overline{\rho {u'w'}}$ and $\overline{\mu{\partial u}/{\partial z}}$, respectively, as is often assumed, because fluctuations in the thermodynamic properties are not negligible.


\subsection{The Role and Mechanism of Density and Viscosity Fluctuations}

The role of $\zeta_{\rho-R}^+$ is evaluated by first considering the turbulent shear stress, $\tau^+_{T-R}$. It is well known that $\tau^+_{T-R}$ results in mixing. When the wall-normal velocity fluctuation, $w'$, is positive, lower momentum from closer to the wall is brought into a higher-momentum region further from the wall. This often causes a negative streamwise velocity fluctuation, resulting in a negative $\overline{u'w'}$, on average.  The opposite occurs for the case of negative $w'$,  and thus the resulting $\tau^+_{T-R}$ is also negative. On the other hand, the role of the density fluctuation term, $\zeta_{\rho-R}^+$, is the opposite, on average. Plotted in Fig.\ref{fig:zeta_rho}, negative $\zeta^+_{\rho-R}$ suggests that the net effect of $\overline{w}\overline{\rho'u'}+\overline{\rho'u'w'}$ is positive in the bulk of the buffer layer and the log layer, and thus $\zeta_{\rho-R}^+$ counteracts the influence of the turbulent mixing. The suggested interpretation is therefore that the fluid inertia opposes the turbulent mixing.\\
\indent While the role of density fluctuations is consistent for all CTBL cases, we find that the role of the viscosity fluctuation terms is different for adiabatic and non-adiabatic cases. For adiabatic cases, Fig.~\ref{fig:zeta_vis} demonstrates that $\zeta^+_{\mu-R}$ is negative, and that the main effect of the viscosity fluctuations is to oppose the turbulent mixing.  For the cold wall cases, $\zeta^+_{\mu-R}$ is positive for  $z^*< 5$, which indicates that $\zeta_{\mu-R}^+$ enhances viscous deceleration, thus counteracting the effect of density fluctuations before dropping to zero in the log layer.\\
\indent Before continuing our analysis, we note that the compressibility influences of pressure fluctuations as well as dilatational velocity fluctuations on hypersonic boundary layer dynamics have also been examined and become a source of great interest in recent years\cite{duan_choudhari_zhang_2016,Yu_Xu_Pirozzoli_2019}. The study of Duan et al. \cite{duan_choudhari_zhang_2016} focuses on the role of the pressure fluctuation in an interaction with freestream acoustic field, for example. And the study of Yu et al.\cite{Yu_Xu_Pirozzoli_2019} examines the effect of dilatational velocity fluctuations on the wall shear stress. Our data suggests that pressure fluctuation magnitudes vary significantly across the current datasets. While linkages undoubtedly exist between pressure, density and viscosity fluctuations, we choose to restrict the current exploration to the influences of thermodynamic variable fluctuations on the near-wall stress balance, leaving additional interesting avenues of study involving pressure for future study.
\subsection{Generalized Favre-Averaged Momentum Balance Equation}
\begin{figure*}
\centering
\begin{subfigure}{.49\textwidth}
  \centering
  \includegraphics[width=\linewidth]{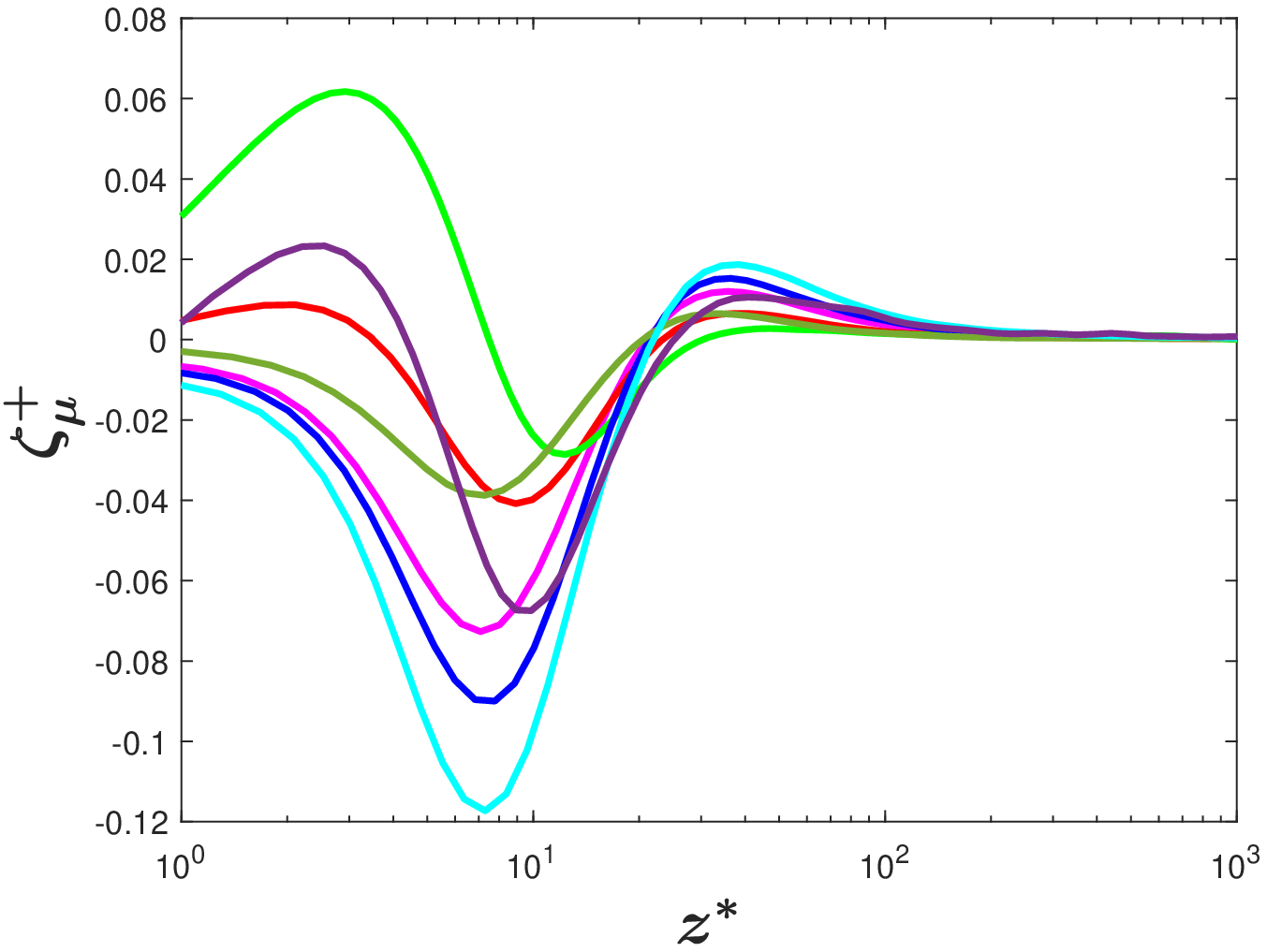}
  \caption{}
  \label{fig:zeta_vis_add}
\end{subfigure}
\begin{subfigure}{.49\textwidth}
  \centering
  \includegraphics[width=\linewidth]{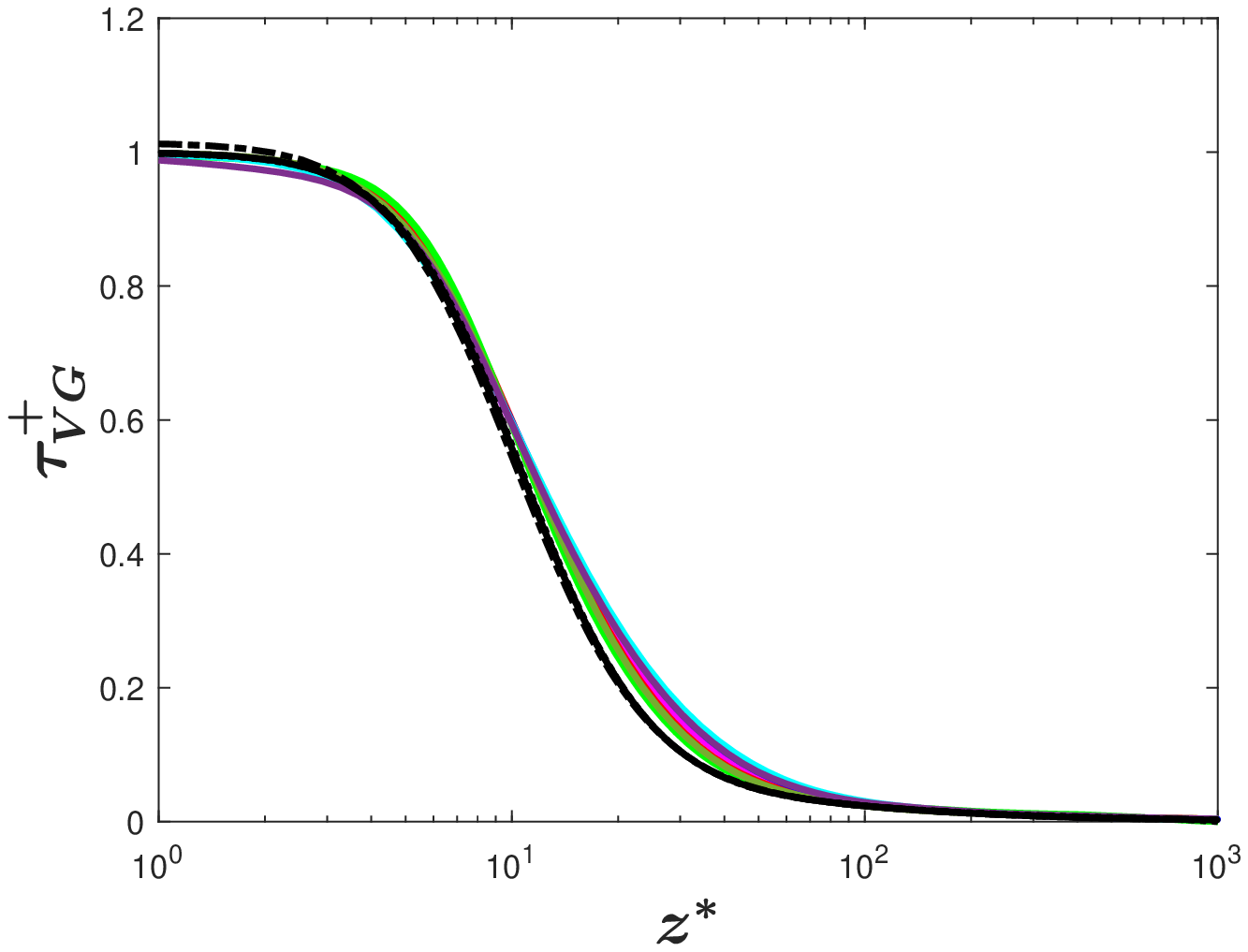}
  \caption{}
  \label{fig:FVtrad_VS}
\end{subfigure}\\%
\begin{subfigure}{.49\textwidth}
  \centering
  \includegraphics[width=\linewidth]{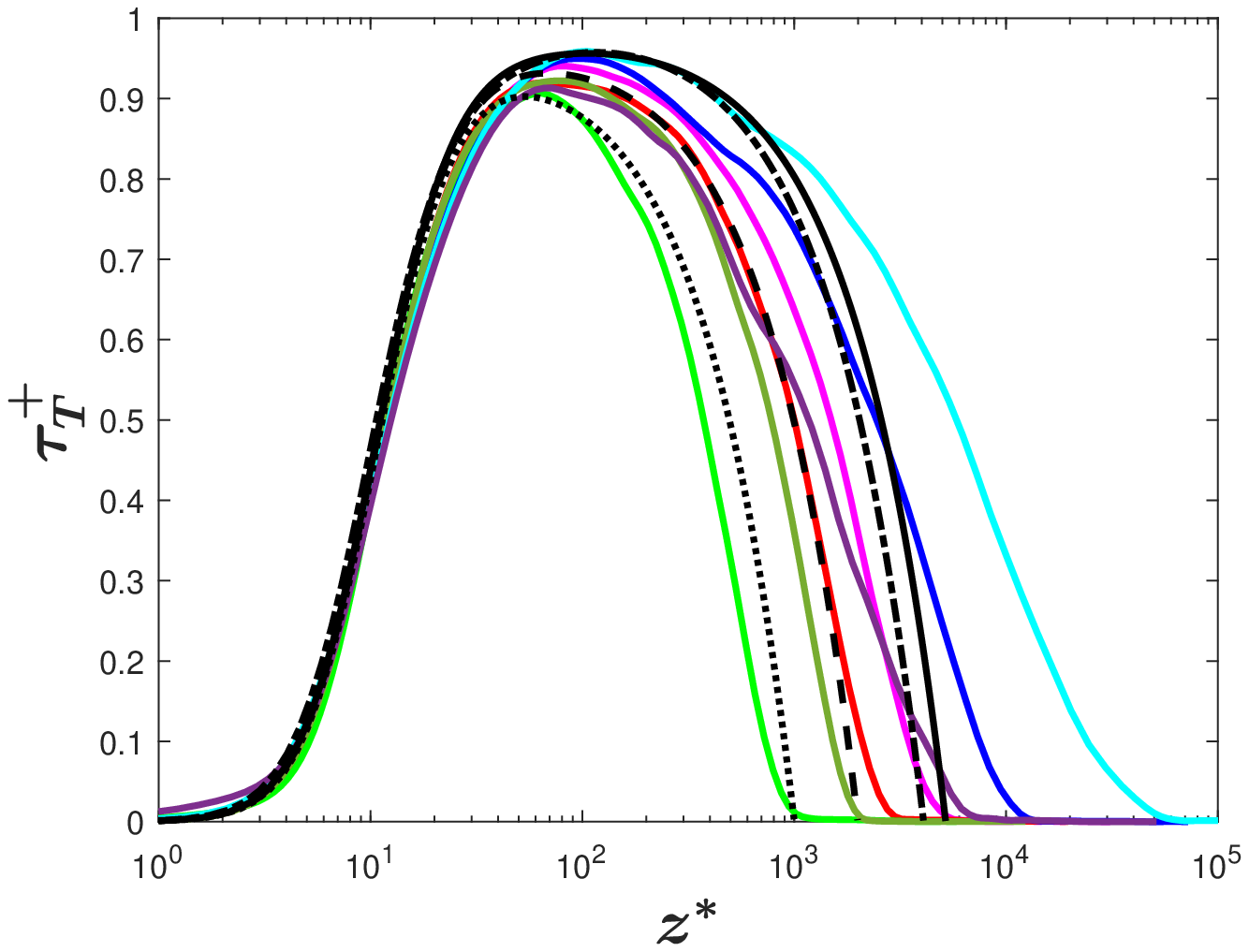}
  \caption{}
  \label{fig:FVtrad_RS}
\end{subfigure}
\begin{subfigure}{.5\textwidth}
  \centering
  \includegraphics[width=\linewidth]{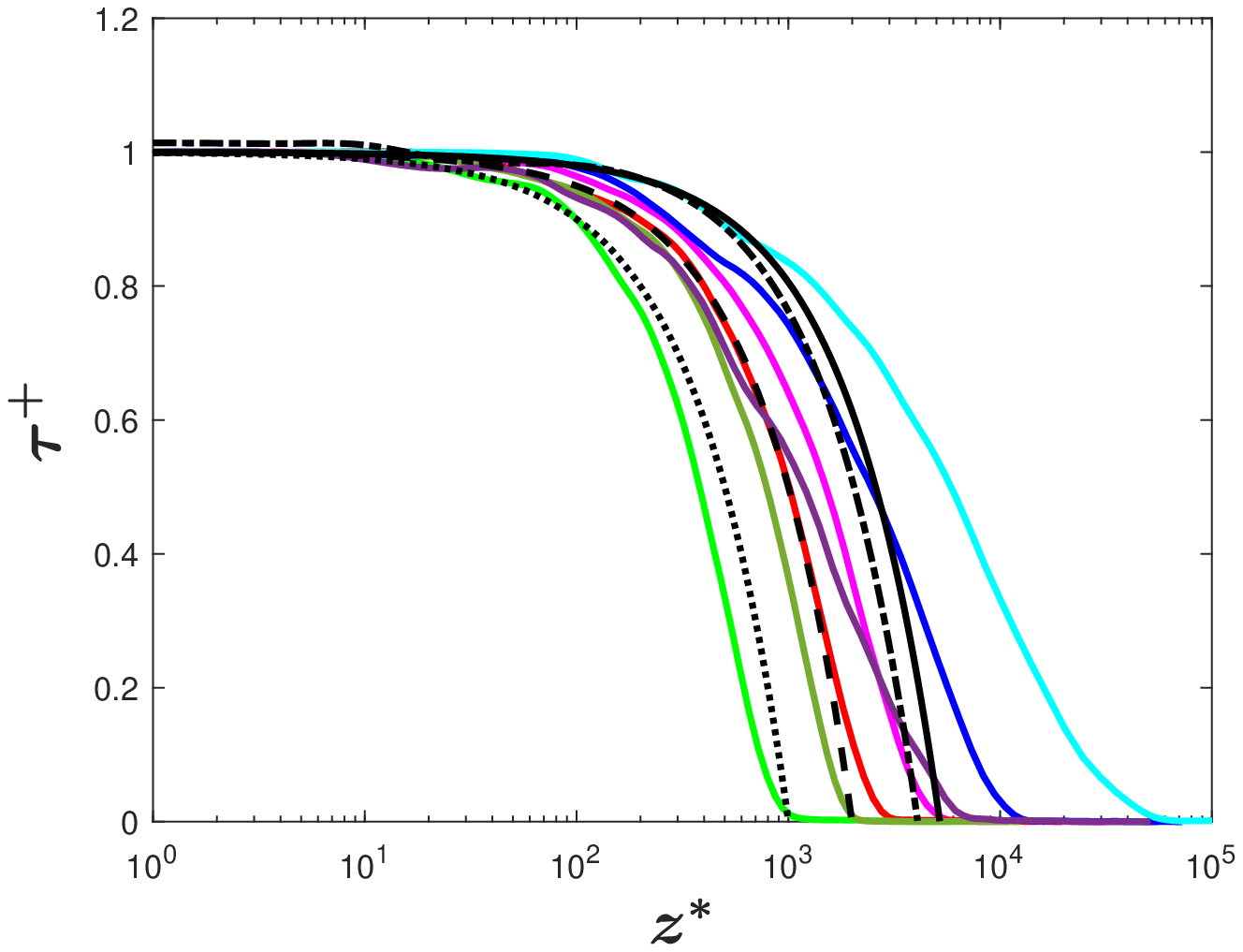}
  \caption{}
  \label{fig:FVtrad_TS}
\end{subfigure}
\caption{Non-dimensional, Favre-averaged (A) viscosity fluctuation terms, $\zeta^+_{\mu}$ (B) viscous stress, $\tau^+_{VG}=\tau_{V}^++\zeta^+_{\mu}$ (C) Turbulent Stress,$\tau_{T}^+$, and (D) total stress,$\tau^+$ are plotted against the semi-local wall normal coordinate, $z^*$. Corresponding stresses from ITBL flow references are plotted for comparison. References for line colors and styles as well as references for the database are included in table \ref{Compressible DNS Database} and \ref{Incompressible DNS Database}.}
\label{fig:FVtradStresses}
\end{figure*}

\indent 
The growing influence of density fluctuations with higher Mach numbers is known but not usually discussed in this context. It is also one reason that Favre averaging has been preferred for compressible flows. It involves scaling a quantity of interest by the instantaneous density (i.e. $\tilde{f} = \overline{\rho f}/\overline{\rho}$). As a result, the Favre-averaged implementation accounts for fluctuations in density since,
\begin{equation}
\label{equation:turbulentstresequaltoeachother}
    \overline{\rho}\widetilde{u''w''} = \overline{\rho u''w''}=\overline{\left(\overline{\rho}+\rho'\right)u''w''}.
\end{equation}
We can therefore write the the near-wall Favre-averaged momentum equation, employing thin shear layer assumptions and  including all fluctuations in thermodynamic properties as  
\begin{equation}
\label{equation:FavreMomentumwithzeta}
    1 = \frac{\left(\overline{\mu}\frac{\partial \tilde{u}}{\partial z} 
    -\overline{\rho}\widetilde{u''w''}\right)}{\tau_w}
    + \left(\underbrace{\frac{\frac{\overline{\overline{\mu}\partial u''}}{\partial z}+\overline{\mu'\frac{\partial u''}{\partial z}}}{\tau_w}}_{\boldsymbol{\zeta_{\mu}^+}}\right)  
    = \left(\underbrace{\tau_{V}^+ + \zeta^+_{\mu}}_{\tau^+_{VG}}\right) +\tau_{T}^+.
\end{equation}
\noindent In this equation, $\overline{\mu}\frac{\partial \tilde{u}}{\partial z}$ is the conventional form of the Favre-averaged viscous stress, $\tau^+_V$, $\overline{\rho}\widetilde{u''w''}$ is the conventional form of the Favre-averaged turbulent stress, $\tau^+_T$, and $\zeta^+_{\mu}$ is the influence arising from viscosity fluctuations. Note that we have employed Reynolds averaging decomposition for the viscosity, as is most common. However, using Favre-averaged decomposition of viscosity, following Spina et al.\cite{SpinaSmitsRobinson}, does not change our conclusions as long as all terms, including terms arising from the viscosity fluctuations, are accounted for. In the following analysis, we define $\tau^+_{VG}$ to be equal to $\tau^+_V+\zeta^+_V$  where the subscript G is used to denote its generality. While  viscosity fluctuations are not normally considered in the conventional Favre-averaged momentum equation due to the use of the Strong Reynolds Analogy\cite{SmitsDussauge}, we demonstrate  that viscosity fluctuations have a non-negligible influence on the near-wall momentum balance in Fig. \ref{fig:zeta_vis_add} and are larger than the corresponding viscous fluctuation term within the Reynolds-averaged equation. \\

\indent The conventional Favre-averaged stress terms, including thermodynamic fluctuations, are plotted in Fig. \ref{fig:FVtradStresses}.
For all cases, the Favre-averaged total stress ,defined as $\tau^+ = \tau^+_{VG} + \tau^+_T$ in Fig. \ref{fig:FVtrad_TS}, remains equal to the wall shear stress (i.e. a value of 1 when nondimensionalized) in the inner layer, deviating at higher $z^*$ for higher $Re^*$, as is the case for ITBLs. Critically, for all CTBL cases considered, we observe a level of Mach-invariance in the total stress, $\tau^+$, that is enforced by the near-wall momentum balance. This result suggests the possible use of the generalized total stress balance for the derivation of a generalized MVT.  The comparison of Favre and Reynolds-averaged turbulent shear stresses ($\tau_{T}^+$ in Fig.~\ref{fig:FVtrad_TS} and $\tau_{T-R}^+$ in Fig.~\ref{fig:Uncorrected_RS})  reveals significant improvements due to the consideration of density fluctuations inherent in the use of Favre-averaging. The resulting turbulent shear stresses are seen to behave quite similarly to the incompressible profiles of Modesti and Pirozzoli \cite{modesti-pirozzoli-2016}. Best comparisons are seen to occur when the compressible $Re^*$ is matched to the incompressible $Re_\tau$ (i.e. M5T1 is compared to LM1000, M3T5, M3Ad, and M5T3; M5T5 is compared to LM2000, M10T3, and M7T5-L; and M12T5-L is compared to LM5200). For all cases, the turbulent stress remains below, but close to, a value of one, consistent with the incompressible theory. 
\indent The magnitude of $\zeta_{\mu}^+$ in Fig.~\ref{fig:zeta_vis_add} indicates that $\tau_{VG}^+$ in Fig.~\ref{fig:FVtrad_VS} would deviate significantly from the wall shear stress and  the ITBL trend for $z* < 5$ if viscosity fluctuations were not taken into account. Moreover, the role of viscosity fluctuations remains unchanged, as discussed in section III-B in relation to $\zeta_{\mu-R}^+$. The increased magnitude of $\zeta_{\mu}^+$, however, suggests that $\zeta_{\mu}^+$ is even more critical to the overall stress balance under the Favre-averaging definition. Interestingly, for both $\zeta_{\mu}^+$ and $\zeta_{\mu-R}^+$, the effect of viscosity fluctuations can remain relevant up to $z^* = 30$ (especially for adiabatic cases) before dropping to smaller values outside of the buffer layer. 
Given the discussion above, it is perhaps not surprising that accounting for density and viscosity fluctuations restores the expected characteristics of the total stress which exhibits good collapse across the ITBL and CTBL data as shown in Fig.\ref{fig:FVtrad_TS}. \\
\indent The stress characteristics discussed above will be used to formulate a new MVT for CTBLs and a few observations important to the following derivation must be reiterated. When all thermodynamic properties important to zero pressure gradient CTBLs are accounted for, namely the mean property gradients as well as density and viscosity fluctuations, the Mach-invariance of the total stress, $\tau^+$ is enforced by the near wall momentum balance. Moreover, for CTBLs, the degree of Mach-invariance for the viscous and turbulent stresses is significantly improved with the new formulations, indicating that the relative contributions from the viscous and turbulent stresses to the total stress remain relatively Mach-invariant in the inner layer. These important observations suggest that there are two types of Mach invariance embedded within $\tau^+$, namely: (1) the Mach-invariance of the generalized near-wall total stress formulation and (2) the Mach-invariance of the relative contributions from the generalized viscous and turbulent stresses to the total stress formulation.

\section{Total Stress Based Velocity Transformation}
\begin{figure*}
\begin{subfigure}{.5\textwidth}
\centering
    \includegraphics[width=\linewidth]{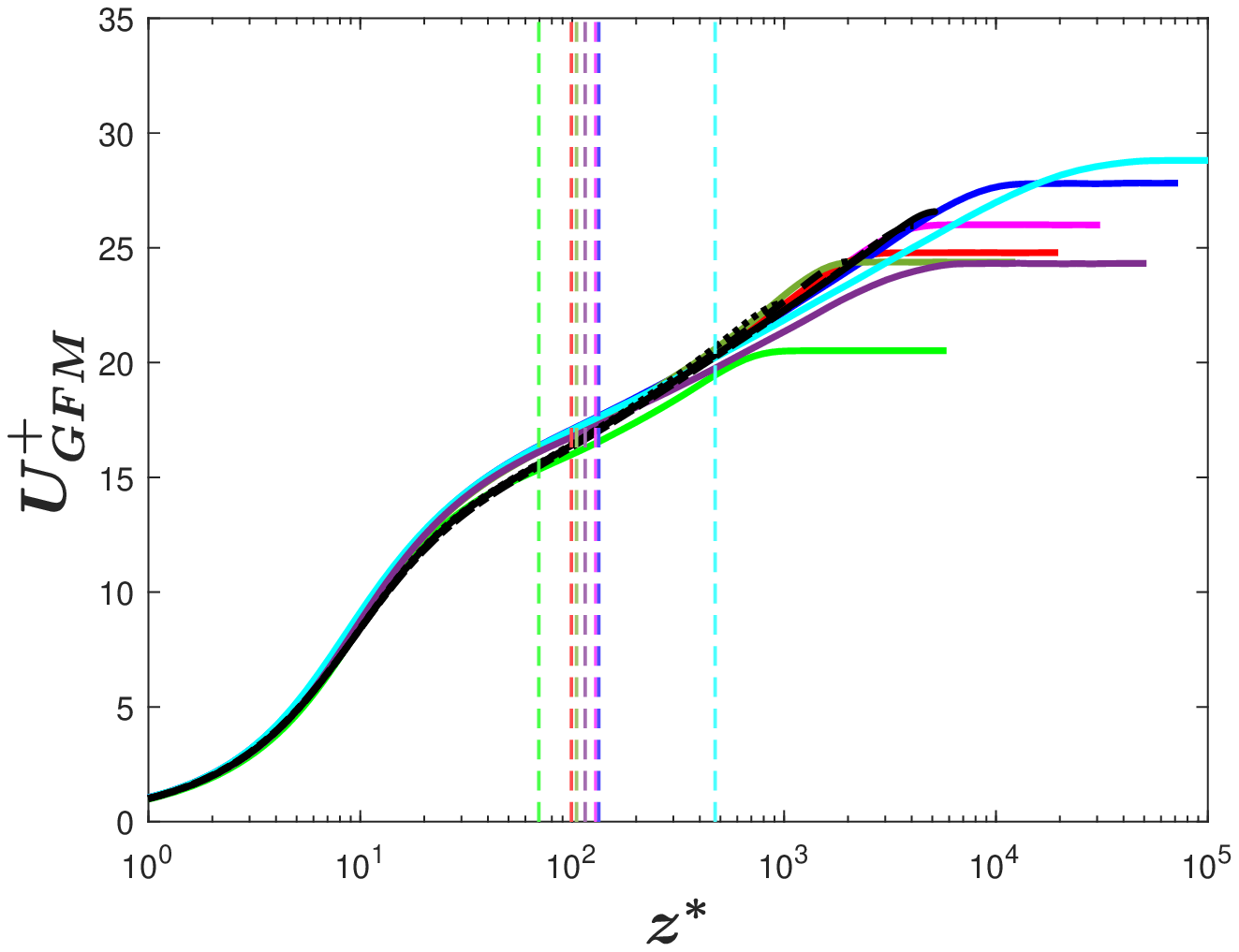}
\caption{}
\label{fig:upls_GFM}
\end{subfigure}
\begin{subfigure}{.5\textwidth}
\centering
\includegraphics[width=\linewidth]{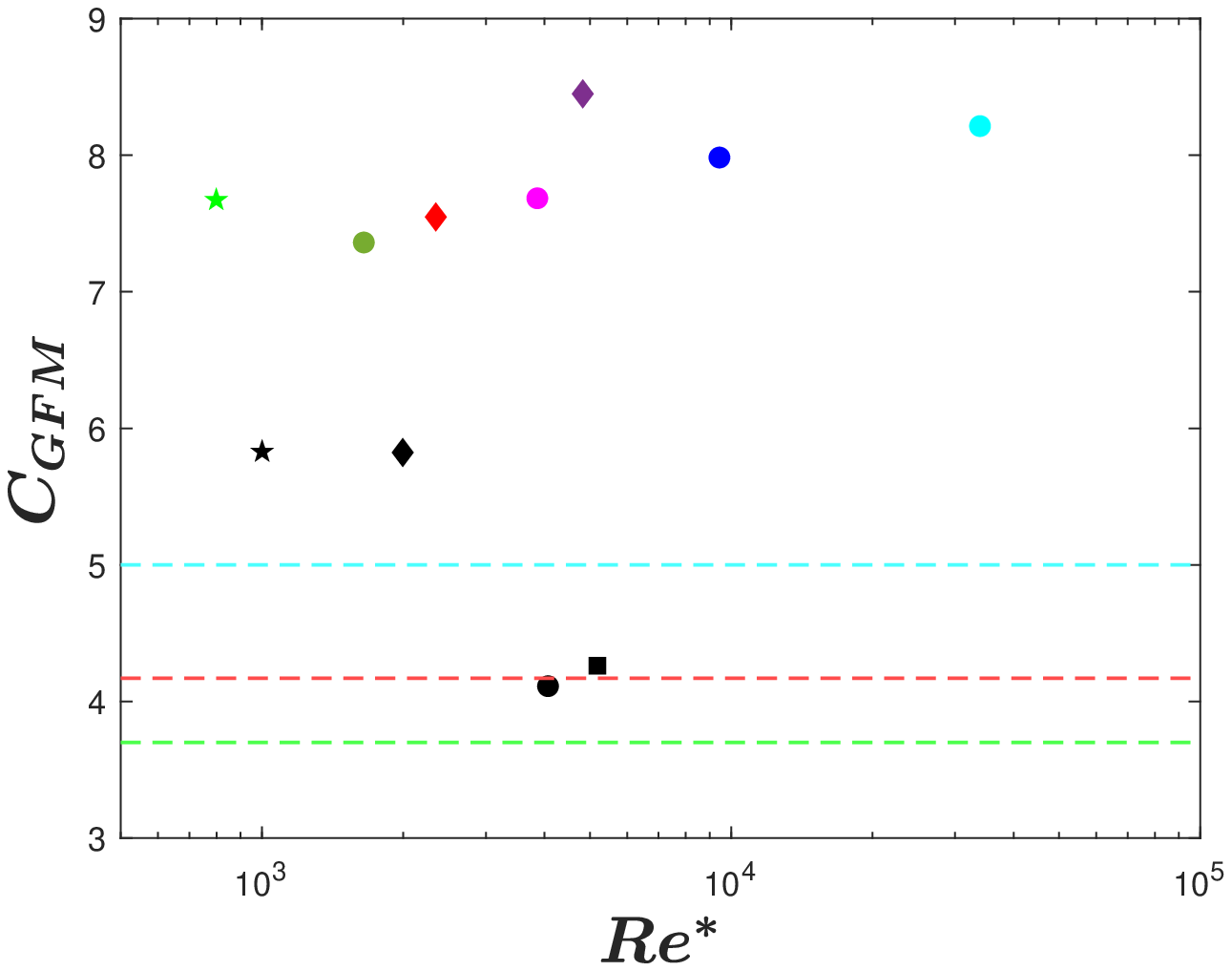}
\caption{}
\label{fig:C_GFM}
\end{subfigure}%
\begin{subfigure}{0.5\textwidth}
\centering
\includegraphics[width=\linewidth]{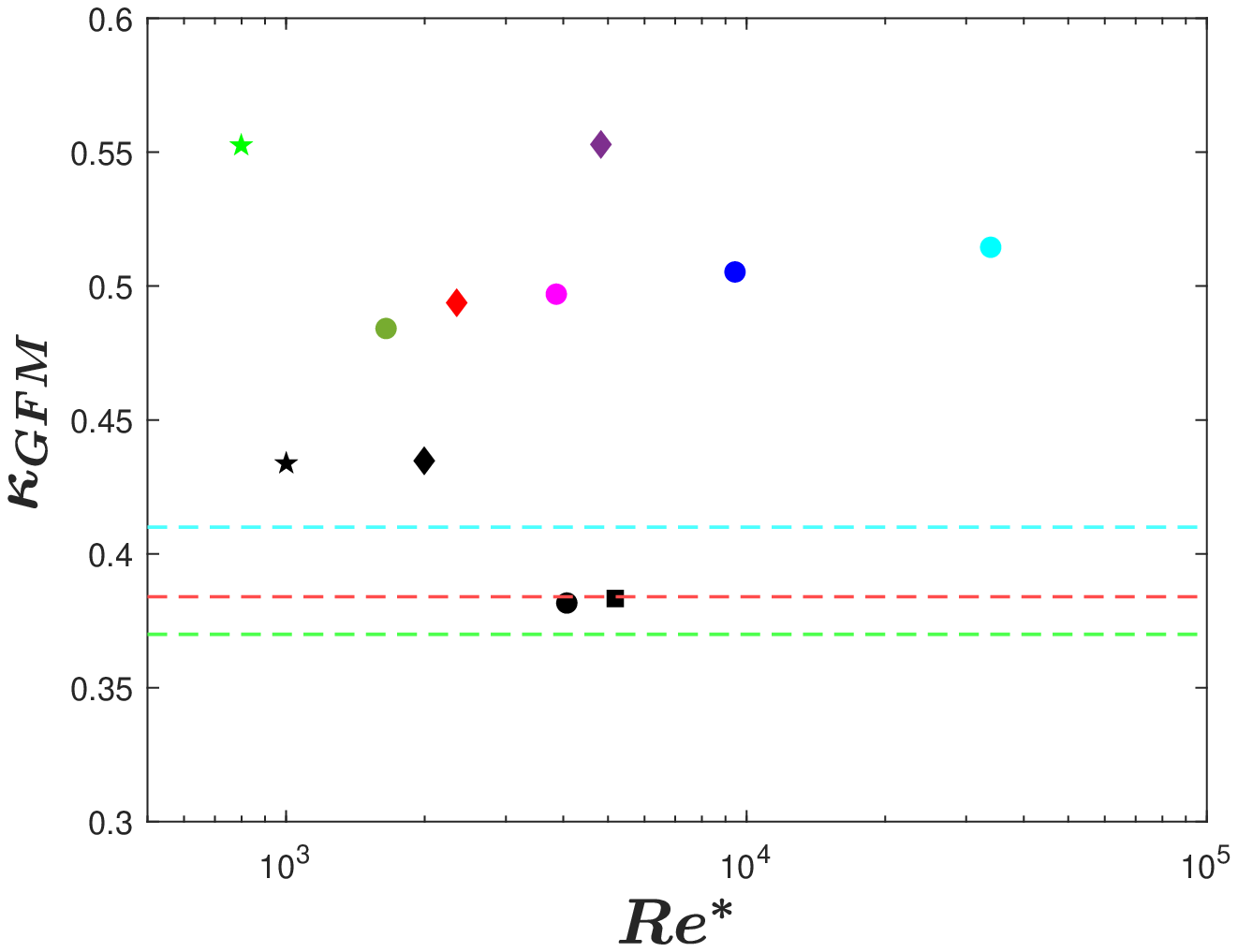}
\caption{}
\label{fig:kappa_GFM}
\end{subfigure}
\caption{ (A) GFM velocity transformation versus semi-local wall normal coordinate, $z^*$. (B) Intercept and (C) Von K\'arm\'an Constant, $\kappa$, of GFM profiles as a function of semi local Reynolds Number, $Re^*$. Classical incompressible law of the wall velocity profiles, intercepts and $\kappa$ of the log layer calculated from the incompressible channel flow database are included for comparison. References for colors and styles of lines and symbols as well as references for the CTBL and ITBL database are included in table \ref{Compressible DNS Database} and \ref{Incompressible DNS Database} unless otherwise noted hereafter. The vertical dashed lines in (A) indicate the wall normal coordinate at which the intercept and $\kappa$ were calculated (Colors match the corresponding CTBL cases with the reference to the colors in table \ref{Compressible DNS Database}). The horizontal dashed lines in (B) and (C) indicate variability in the intercept and $\kappa$, respectively, reported in \cite{Nagib_Chauhan_2008}, [dashed cyan line-superpipe, dashed red line- boundary layer, dashed green line-channel]. }
\label{fig:GFM}
\end{figure*}
\begin{figure}
    \centering
    \includegraphics[width=.5\linewidth]{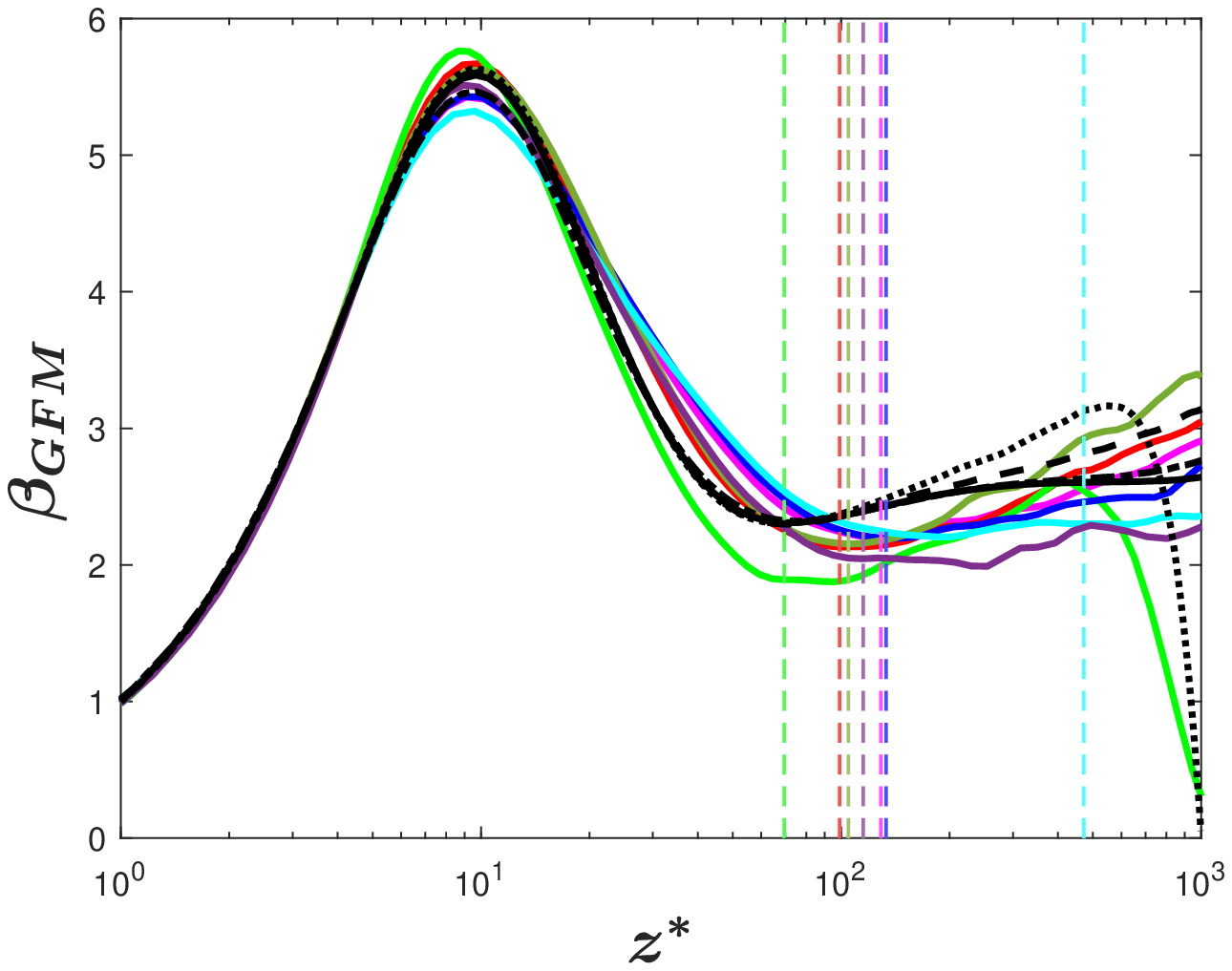}
    \caption{Pre-multiplied mean shear, $\beta_{GFM}$, based on GFM transformation versus semi-local wall normal coordinate. Pre-multiplied mean shear of LM5200 and BOP4100 is calculated from the classical incompressible law of the wall velocity profile and plotted as incompressible reference for comparison. The references for line colors and styles as well as the references for the database are included in table \ref{Compressible DNS Database} and \ref{Incompressible DNS Database} unless otherwise noted hereafter. The vertical dashed lines indicate the wall normal coordinate at which the log layer parameters were calculated (Colors match the corresponding CTBL cases with the reference to the colors in table \ref{Compressible DNS Database}). The dashed grey lines are publicly available CTBL database of Zhang et al.~\cite{ZhangLian2018}. }
    \label{fig:beta_GFM}
\end{figure}
In recent months, a new MVT based on total stress has been proposed by Griffin, Fu \& Moin~\cite{GriffinMoin}. Mean velocity profiles based on their transformation are shown in Fig.~\ref{fig:upls_GFM}, exhibiting a good collapse to the incompressible law of the wall. Success of this transformation is attributed to the use of a total-stress-based balance to combine the viscous stress-based transformation by Trettel \& Larsson~\cite{TL} and the quasi-equilibrium-based transformation of Zhang et al.~\cite{ZhangHussain} at locations where the assumptions underlying each transformation are valid. The  mathematical form of the total-stress-based transformation can be derived from the total stress representation of the mean shear, resulting in, 
\begin{equation}
    \label{equation:totalrepresentation_of_meanshear}
    \tau^+=S_t^+\left(\frac{\tau_V^+}{S_{TL}^+}+\frac{\tau_{T}^+}{S^+_{eq}}\right)
\end{equation}
\noindent where $S_{TL}^+= \overline{\mu}^+\partial \tilde{u}^+/\partial z^+$ denotes non-dimensinalized mean shear transformed according  to Trettel \& Larsson~\cite{TL} and $S_{eq}^+=({1}/{\overline{\mu}^+}){\partial \Tilde{u}^+}/{\partial z^*}$ denotes the non-dimensionalized mean shear of Zhang et al.~\cite{ZhangHussain} generalized by Griffin, Fu \& Moin \cite{GriffinMoin} to semilocal wall units, $z^*$. This result can then be rearranged to obtain, 
\begin{equation}
    \label{equation:Stpls}
    S_t^+=\frac{\tau_T^+S_{eq}^+}{\tau^++S_{eq}^+-S_{TL}^+}.
\end{equation}
\noindent Once calculated, the resulting transformed, non-dimensional shear stress, $S_t^+ = \partial U_{GFM}^+/\partial z^*$, can the be integrated to obtain the transformed velocity velocity profile, $U_{GFM}^+$.  Note that they employ Favre averaged variables, accounting for density  but not viscous fluctuations.

\indent In the study of Griffin et al.~\cite{GriffinMoin}, they utilize a CTBL database with $Re^*$ approximately ranging from 200 to 4900 and Mach number ranging from 2 to 14. They present the integrated error of the transformed mean velocity profile relative to the conventional incompressible log-law. The database of the  current paper allows us to test the GFM transformation over a wider range of $Re^*$, including a range of diabatic conditions (see table \ref{Compressible DNS Database}). Following Griffin et al.~\cite{GriffinMoin}, comparisons are made with the incompressible cases of Lee \& Moser (2015)\cite{LeeMoser15} and Bernardini and Pirozzoli (2014)\cite{BOP} (see Table \ref{Incompressible DNS Database}). The  LM5200 ITBL was used by Griffin et al.~\cite{GriffinMoin} for their incompressible law of the wall baseline. \\
\indent In contrast to Griffin et al. we choose to examine the influence of the transformation on the variation in slope and intercept of the log layer. These parameters are determined using the pre-multiplied mean shear as suggested by Lee \& Moser~\cite{LeeMoser15}, given by $\beta = z^* dU_t^+/dz^*$, where $U_t^+$ is a transformed mean velocity of interest. If there is a logarithmic layer, the pre-multiplied mean shear will have a plateau, or constant-valued region. To determine the location of the plateau (or best approximation to a plateau should a plateau not fully exist) $d\beta/dz^*$ is calculated from $z^*=30$ to  $z/\delta=0.2$, where $\delta$ is the wall normal coordinate at the location where mean velocity is 99\% of the freestream value. The $z^*$ location at the minimum value of  $d\beta/dz^*$ is used as the ``characteristic location'' for the log layer and the location at which a slope and intercept are calculated. This location is thus not dependent  on the chosen bounds of the search. \\
\indent The log layer intercept and slope for the current CTBL database under the GFM transformation are shown in Fig.~\ref{fig:C_GFM} and Fig.~\ref{fig:kappa_GFM}, where $C$ is the intercept and $\kappa$ is 1/slope or the K\'arm\'an constant. The calculated ``characteristic locations'' of the logarithmic  portion of each profile are also shown in Fig. \ref{fig:upls_GFM}. Also shown are reference ITBL log-law characteristics and the variability of ITBL slope and intercept reported in the literature by Nagib \& Chauhan \cite{Nagib_Chauhan_2008}. The slope and intercept derived from the GFM transformation of the current compressible datasets lie outside the range of values commonly reported for ITBLs for higher Reynolds numbers. Both $\kappa$ and intercept are seen to be larger under this transformation. The slope and intercept are also larger than the low Reynolds number incompressible cases of cases of Lee and Moser  \cite{LeeMoser15}. In an attempt to improve the GFM transformation results, we have tested a version of the transformation where $\tau^+_{VG}$ was used to replace $\tau^+_{V}$, in an effort to account for the influence of viscosity fluctuations on  the near-wall stress balance. This test did not show an appreciable improvement to slope or intercept values and thus viscous fluctuations  are not the source of the observed disparity between the incompressible log-law and GFM transformed results. A further examination of the pre-multiplied mean shear, which is plotted in Fig.~\ref{fig:beta_GFM}, suggests that the collapse for the GFM mean velocity profile in Fig.~\ref{fig:upls_GFM} is observed because Mach-invariance of the pre-multiplied mean shear is somewhat satisfactory up to $z^* = 20$ in the middle of the buffer layer. However, the Mach-invariance of $\beta_{GFM}$ quickly deteriorates in the region at $z^* > 20$ where the quasi-equilibrium model starts to represent the turbulent stress. This incomplete Mach-invariance extends to the plateau region, where it is expected to show the most logarithmic behavior, resulting in incorrect intercepts and slopes within the log layer region. This observation, which points to the log layer as the source of the error, is perhaps an explanation as to why the use of $\tau^+_{VG}$  showed  negligible improvement. While higher $Re^*$ cases show larger regions of logarithmic behavior, it is unclear if the correspondence of slope and intercept with incompressible data will improve significantly for the CTBLs at some combination of high $Re^*$ or $Re_\tau$. Despite this, we should note that  the GFM transformation does provide a better collapse of the compressible profiles than most previously proposed MVTs.  \\ 
\begin{figure*}
\centering
    \includegraphics[width=.5\linewidth]{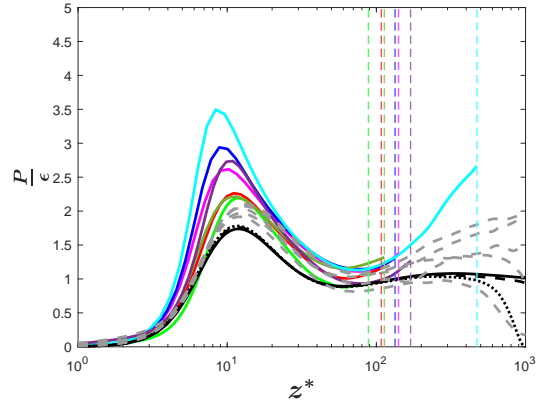}
    \caption{ The ratio of the TKE production (P) and the viscous dissipation ($\epsilon$) versus semi-local wall normal coordinate, $z^*$. The vertical dashed lines indicate the wall-normal coordinate at which the log layer parameters were calculated. Grey dashed lines are CTBL data from Zhang et al.\cite{ZhangLian2018}. References for line colors and styles as well as the references for the database are included in table \ref{Compressible DNS Database} unless otherwise noted.}
    \label{fig:P_epsilon_ratio_alpha}

\end{figure*}
\indent We further examine the reasons for the above variability of slope and intercept by exploring the validity of the assumptions undertaken in the derivation of the transformation by Zhang et al.~\cite{ZhangHussain}, namely the Mach-invariance of (1) the Favre-averaged enstrophy, (2) the near-wall turbulent stress profile, and (3) of the ratio of the Favre-averaged turbulent kinetic energy (TKE) production and viscous dissipation terms. For brevity in this section, Einstein notation is used to express the terms pertaining to the TKE equation, where the definition of $v$ is the velocity in the direction determined by the indices, $i$,$j$ and $k$. \\
\indent While not shown here, it is confirmed that the Mach-invariance of the Favre-averaged enstrophy suggested  by Lagha et al. \cite{lagha-kim-zhong-2011} holds for the current CTBL cases. Recall also that the turbulent stress in Fig.~\ref{fig:FVtrad_RS} is Mach-invariant in the log layer when employing Favre averaging. Finally, we explore the ratio of the Favre-averaged TKE production (P) and the Favre-averaged viscous dissipation $(\epsilon)$,  mathematically given by 
\begin{equation}
    \label{equation:TKE_P}
    P=\overline{\rho}\widetilde{v_{i}''v_{j}''}\frac{\partial \widetilde{v_{i}}}{\partial x_{j}}
\end{equation}

\begin{equation}
\begin{split}
    \label{equation:TKE_e}
    \epsilon = \overline{{\mu}\left(\frac{\partial (\tilde{v}+v'')_i}{\partial x_j} + \frac{\partial (\tilde{v}+v'')_j}{\partial x_i} -\frac{2}{3}\delta_{ij}\frac{\partial (\tilde{v}+v'')_k}{\partial x_k}\right) \frac{\partial v''_i}{\partial x_j}}.
\end{split}    
\end{equation}

Similar to the Favre-averaged turbulent stress, the Favre averaged TKE production term  accounts for density fluctuations by Favre-averaging definition, and the TKE dissipation is defined to include viscosity fluctuations by explicitly using the instantaneous viscosity. 
The ratio of the TKE production (P) and viscous dissipation ($\epsilon)$ terms is plotted in Fig.~\ref{fig:P_epsilon_ratio_alpha}.   The Mach-invariance deteriorates in the viscous and buffer layers where it is not expected hold\cite{ZhangHussain}, while the Mach-invariance improves in the log layer where collapse is more expected. Despite this, an observation of the $P/\epsilon$ ratio from our CTBL datasets as well as the CTBL data from Zhang et al. \cite{ZhangLian2018} suggests that it remains unclear whether the quasi-equilibrium assumption exhibits sufficient Mach-invariance to form the basis for a velocity transformation in the log layer region.

\section{Proposed new total stress-based transformation}
\begin{figure}
    \centering
    \includegraphics[width=.5\linewidth]{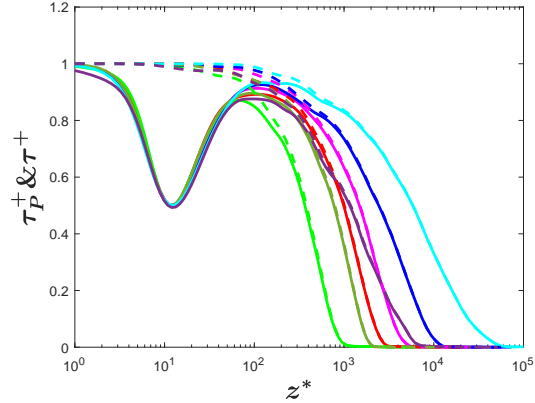}
    \caption{Solid line indicates proportionally accurate generalized total stress,$\tau^+_{P}$, and dashed line indicates generalized total stress, $\tau^+$ plotted for comparison. References for lines colors as well as references for the database are included in table \ref{Compressible DNS Database}.}
    \label{fig:tau_TS_alpha_p}
\end{figure}

In an effort to develop a generalized MVT for CTBL, we revisit the total stress representation in Eq.~\ref{equation:totalrepresentation_of_meanshear} by Griffin, Fu \& Moin \cite{GriffinMoin}. They represent the total stress in terms of mean shear quantities, $S_t^+,S_{TL}^+$ and $S_{eq}^+$.  By construction, $S_t^+$ will exhibit the characteristics of either $S_{TL}^+$ in the near-wall limit, or of $S_{eq}^+$ in the log layer. More importantly, Eq. \ref{equation:totalrepresentation_of_meanshear} preserves the magnitude of the total stress prescribed by the viscous and the turbulent stresses. Therefore, we propose several modifications to Eq.~\ref{equation:totalrepresentation_of_meanshear} to utilize the scaling properties of $\tau^+$ identified earlier: (1) the near-wall Mach-invariance of the $\tau^+$ magnitude dictated by the momentum balance and (2) the Mach-invariance of the relative contributions from the generalized viscous and turbulent stresses to $\tau^+$ . \\
\indent The validity of the first property was demonstrated by Figure \ref{fig:FVtrad_TS}. Therefore, both viscous stress and turbulent stress in Eq.~\ref{equation:totalrepresentation_of_meanshear} are replaced with $\tau_{VG}^+$ and $\tau_{T}^+$, respectively as shown in Eq.~\ref{equation:density_corrected_total_stress_by_mean_shear}. Also, note that the mean shear form of viscous stress, $S_{V}^+$, is the same as the viscous stress such that $S_{V}^+ = \tau_{VG}^+$ and we make use of this simplification in a similar manner to the GFM transformation. The second scaling property of $\tau^+$ is enforced by replacing $S^+_{eq}$ by $S^+_{P}$, the details of which will be discussed subsequently. The resulting equation is shown in Eq.~\ref{equation:density_corrected_total_stress_by_mean_shear} where the generalized mean shear, $S_{G}^+$, can be integrated with respect to the semilocal wall unit, $z^*$, from the wall to the freestream to obtain the transformed velocity, $U_{G}^+=\int S_{G}^+ dz^*$.
\begin{equation}
    \label{equation:density_corrected_total_stress_by_mean_shear} 
    \tau^+=S_{G}^+\left(\frac{\tau_{VG}^+}{S_{V}^+}+\frac{\tau_{T}^+}{S^+_{P}}\right) = S_{G}^+\left(1+\frac{\tau_{T}^+}{S^+_{P}}\right)
\end{equation}
Before solving for $S_{G}^+$, an expression must be found for for $S^+_{P}$ which includes the Mach-invariance of the relative contributions from the generalized viscous and turbulent stresses. We begin by defining $R_{V}$ and $R_{T}$ to be the ratio of the viscous and turbulent stresses to the total stress respectively.
\begin{equation}
    \label{equation:RvsRrs}
    \begin{split}
     R_{V}&=\tau^+_{VG}/\tau^+\\
     R_{T}&=\tau^+_{T}/\tau^+      
    \end{split}
\end{equation}
The Mach-invariance in the relative contributions of $\tau_{VG}^+$ and $\tau_{T}^+$ to $\tau^+$ is first ensured by multiplying $R_{V}$ and $R_{T}$ by the viscous and turbulent stress terms, 
\begin{equation}
    \label{equation:proportionaltotalstress}
    \begin{split}
        \tau_{V,P}^+ = R_{V}{\tau}^+_{VG} \\
        \tau_{T,P}^+ = R_{T}{{\tau}_{T}^+}
    \end{split}
\end{equation}
\noindent where $\tau^+_{V, P}$ and $\tau^+_{T, P}$ provide an accurate proportional representation of each stress to the total stress at any wall-normal coordinate location. This mathematical treatment decouples the viscous friction, $\tau_{VG}^+$, and turbulent mixing, $\tau_{T}^+$, effects. Also, note in Fig.~\ref{fig:tau_TS_alpha_p} that $\tau^+_{P}$, defined as $\tau^+_{P} = \tau_{V,P}^+ + \tau_{T,P}^+$, is seen to remain close to Mach-invariant in the near-wall region, as designed. The proportional total stress,  $\tau^+_{P}$ is seen to vary with Mach number in the outer layer region, as its value is reduced after its second peak. Also plotted here is the generalized form of the total stress, $\tau^+$, for comparison. \\
\indent Provided with an accurate proportional representation of each stress from Eq.~\ref{equation:proportionaltotalstress}, we can now  define a proportional mean shear, $S^+_{P}$. We start by assuming that law of the wall for compressible turbulent boundary layer exists, ie, $dU_{G}^+/dz^* = \tau_{VG}^+$ in the viscous layer and $dU_{G}^+/dz^* = \sqrt{\tau_{T}^+}/kz^*$ in the turbulent layer. It might be assumed that the sum of the viscous and log-layer shear forms would describe the mean shear at a given wall-normal location. However, to ingrain the stress-proportionality property into the mean shear form, the viscous and turbulent stresses are replaced by $\tau_{V,P}^+ $ and $\tau_{T,P}^+$, resulting in  equation~\ref{equation:proportional_total_stress_mean_shear}. 
\begin{equation}
    \label{equation:proportional_total_stress_mean_shear}
    S^+_{P}=R_{V}\boldsymbol{\tau}^+_{VG}+\frac{\sqrt{R_{T}({\boldsymbol{\tau}^+_{T})}}}{\kappa z^*},     
\end{equation}
Note that in the respective layer where each stress dominates, the value of the each stress term approaches 1, thus restoring a similar incompressible law of the wall form, $d\overline{u}^+/dz^+ = 1$ or $\overline{\mu_w} {d\overline{u}/{dz}} = \tau_w$ in the viscous layer and $d\overline{u}^+/dz^+ = 1/kz^+$ or $d\overline{u}/dz = {\sqrt{\tau_w/\overline{\rho}_w}}/{kz}$ in the turbulent layer. \\
\indent While the mathematical treatment of the second term in Eq.~\ref{equation:proportional_total_stress_mean_shear} is the same as that of the mixing length hypothesis in that the turbulent stress is also square-rooted and divided by the mixing length, $\kappa z^*$, the present paper does not endorse the mixing length hypothesis. Rather, this mathematical treatment was derived based on the assumption that law of the wall exists for compressible turbulent boundary layer. For the velocity transformation and for all cases in this paper, we use the value  of 0.381  for $\kappa$ from LM5200 in table \ref{Incompressible DNS Database}, as reported in Lee \& Moser\cite{LeeMoser15}.\\
\indent Finally, with $S^+_{P}$ derived, which provides the information regarding the proportional contribution of the stresses to the total stress, equation \ref{equation:density_corrected_total_stress_by_mean_shear} can now be rearranged to solve for the mean shear, $S_{G}^+$, enforcing the correct magnitude of the total stress such that
\begin{equation}
    \label{equation:final_V_total_stress_mean_shear} 
    S_{G}^+=\frac{\tau^+}{1+\tau_{T}^+/S^+_{P}}.
\end{equation}
\indent Thus, the mean shear, $S_{G}^+$, in Eq.\ref{equation:final_V_total_stress_mean_shear} provides a newly proposed transformation that preserves the two scaling properties of $\tau^+$ as described in the beginning of the section.  Moreover, by the definition of each constituting term and by construction, this transformation includes the effects of mean density and viscosity variation, density and viscosity fluctuations, and viscous and turbulent stress balances. \\

\section{Discussion of Results}
\begin{figure*}
    \centering
\begin{subfigure}{.5\textwidth}
    \centering
    \includegraphics[width=\linewidth]{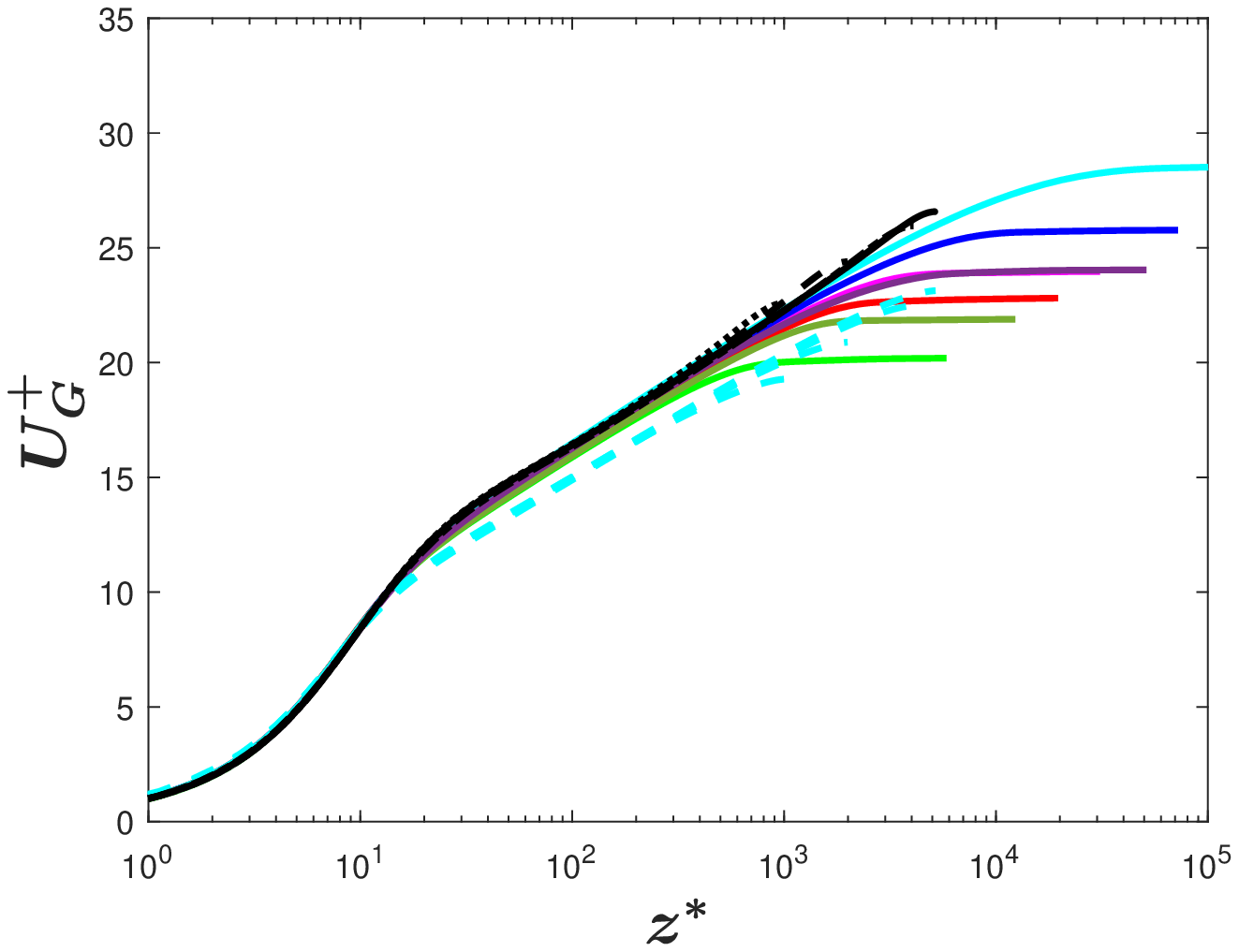}
    \caption{}
    \label{fig:upls_Han}  
\end{subfigure}%
\begin{subfigure}{.5\textwidth}
    \centering
    \includegraphics[width=\linewidth]{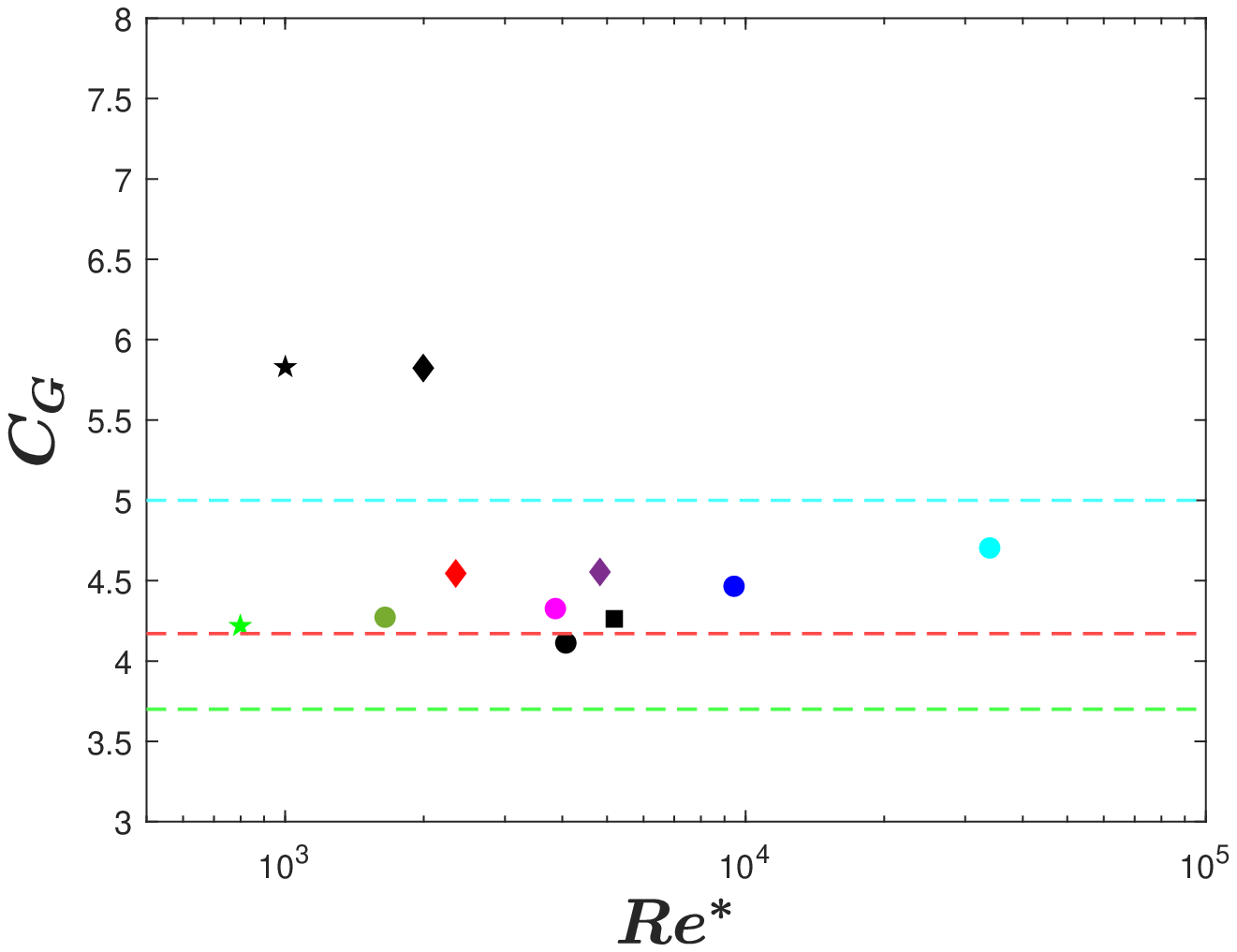}
    \caption{}
    \label{fig:C_Han}  
\end{subfigure}
\begin{subfigure}{.5\textwidth}
    \centering
    \includegraphics[width=\linewidth]{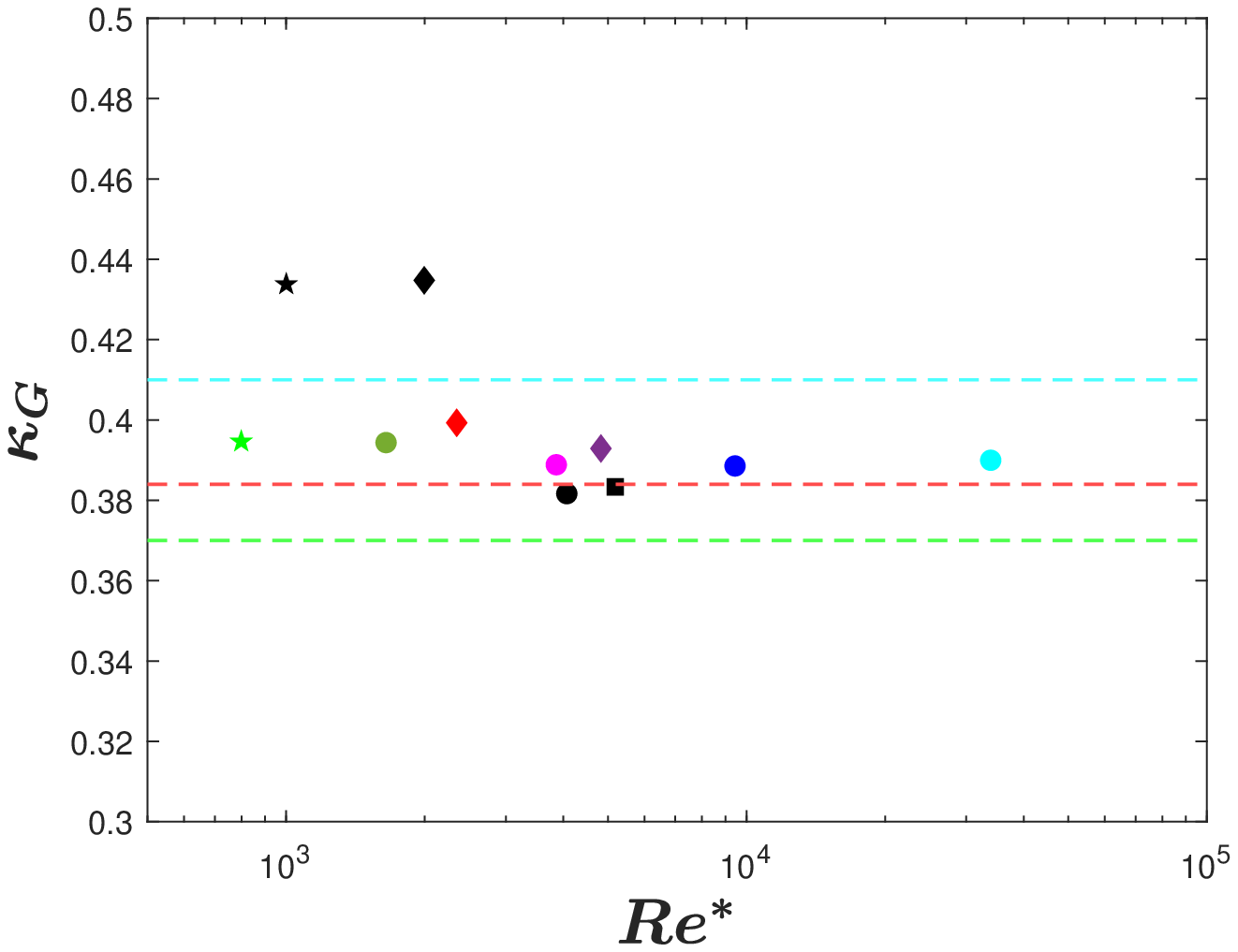}
    \caption{}
    \label{fig:kappa_Han}  
\end{subfigure}
\caption{(A) The presently-proposed velocity transformation versus semi-local wall normal coordinate, $z^*$. (B) Intercept and (C) von K\'arm\'an constant, $\kappa$, of the log layer plotted against semi-local Reynolds number. The ITBL flow data listed in table \ref{Incompressible DNS Database} are transformed using a conventional law of the wall and are plotted for comparison in (A),(B) and (C).  References for line colors and styles and symbols as well as references for the database are included in table \ref{Compressible DNS Database} and \ref{Incompressible DNS Database} unless noted otherwise hereafter. Dashed cyan lines in (A) denote the velocity profiles of the ITBL flow data transformed by the proposed MVT. Horizontal lines in (B) and (C) denote variability in the intercept and $\kappa$, respectively, reported in Nagib and Chauhan\cite{Nagib_Chauhan_2008}, (dashed cyan line, superpipe), (dashed red line, boundary layer) and (dashed green line, channel). }
\label{fig:Han}
\end{figure*}
\begin{figure}
    \centering
    \includegraphics[width=.5\linewidth]{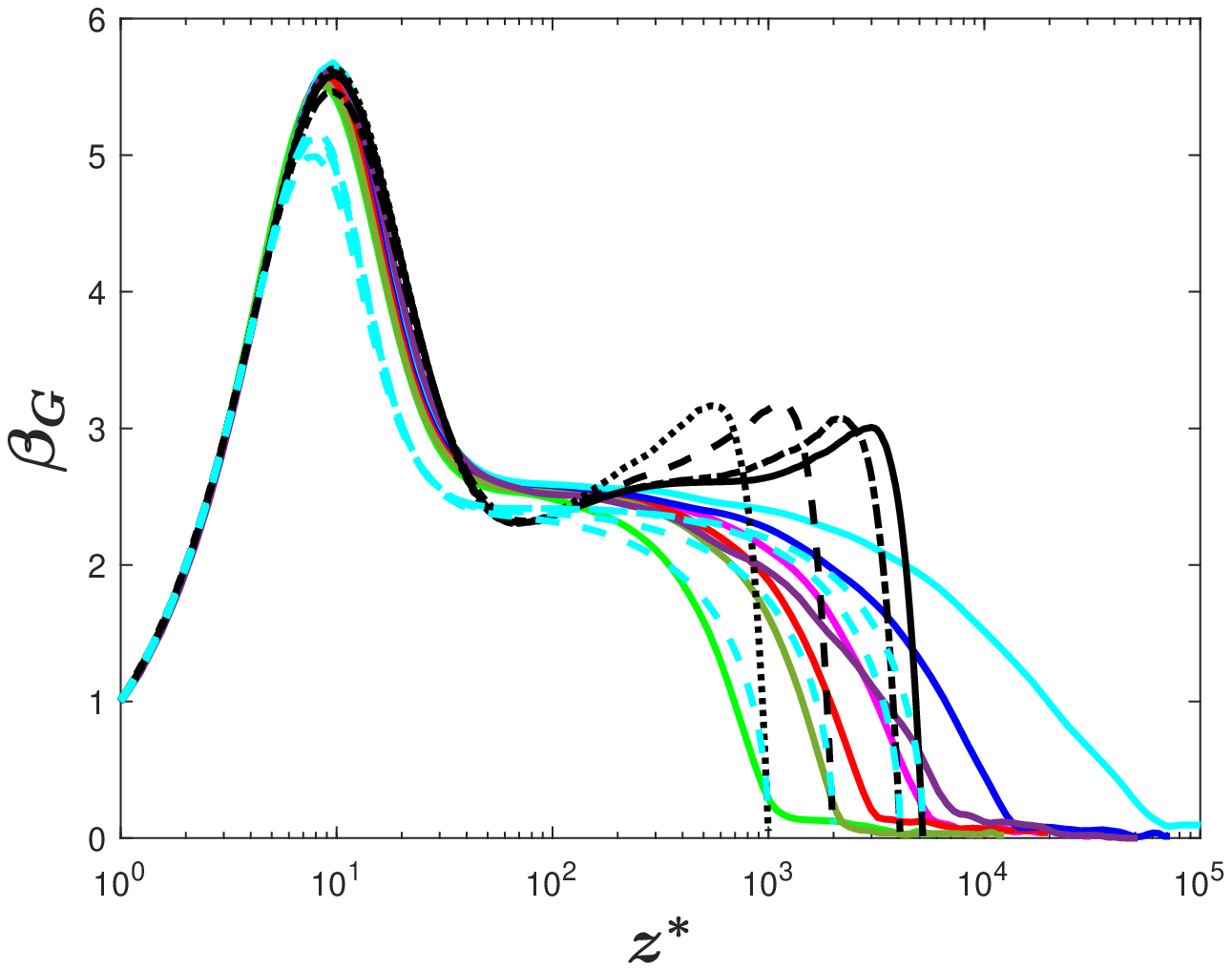}
    \caption{Pre-multiplied mean shear, $\beta_{G}$,based on the present mean velocity transformation versus semi-local wall normal coordinate, $z^*$. The pre-multiplied mean shear is calculated for ITBL cases (see table \ref{Incompressible DNS Database}) using the conventional law of the wall for comparison. References for line colors and styles as well as references for the database are included in table \ref{Compressible DNS Database} and \ref{Incompressible DNS Database} unless noted otherwise hereafter. Dashed cyan lines indicate the pre-multiplied mean shear of the ITBL cases transformed by the proposed MVT. }
    \label{fig:beta_alpha}
\end{figure}
\indent The transformed mean velocity profiles for CTBL according to Eq. \ref{equation:final_V_total_stress_mean_shear} are shown in Fig.~\ref{fig:upls_Han}. Also included here is ITBL data that follows the classical law of the wall. The proposed transformation is seen to collapse all of the CBTL velocity profiles (supersonic and hypersonic, adiabatic and non-adiabatic) to the classical incompressible result. In addition to the qualitative examination of the mean velocity profile, both the log-law intercept and slope for all CTBL and ITBL data are extracted in a similar manner to the earlier analysis of the GFM transformation. The characteristic location of the log layer was found  using the first order derivative of the pre-multiplied mean shear profile ($\beta_G = z^* dU_{G}^+/dz^{*}$, see Fig.~\ref{fig:beta_alpha}). Note that slope and the intercept of the ITBL data in Fig.~\ref{fig:C_Han} and Fig.~\ref{fig:kappa_Han} are calculated from the classical law of the wall profiles. With an exception of low $Re_{\tau}$ incompressible cases, which exhibit an expected Reynolds number dependency, the slope and the intercept for CTBL cases in Fig.~\ref{fig:C_Han} and Fig.~\ref{fig:kappa_Han} show a very small variability when compared to highest Reynolds  number ITBL cases (LM5200\cite{LeeMoser15} and BOP4200\cite{BOP}). The slope and intercept of the CTBL datasets  all lie within the bounds of high Reynolds number values seen for incompressible flow \cite{Nagib_Chauhan_2008}.  The proposed transformation accounts for the relative contributions of viscous and turbulent stresses directly, thus it also accounts for one of the main sources of low-Reynolds number dependence on the classical law of the wall.\\ 
\indent To quantitatively compare the scatter in the log law intercept and slope for the proposed and GFM transformations, we use the coefficient of variation (CoV), which is defined as the standard deviation divided by the mean of the data of interest.  Higher CoV values indicate larger scatter.  The intercept and $\kappa$ value for LM5200\cite{LeeMoser15} and BOP 4100\cite{BOP} are also considered to measure the scatter with a reference to ITBL cases.  Incompressible cases with lower Reynolds numbers, namely LM1000\cite{LeeMoser15} and LM2000\cite{LeeMoser15}, are excluded as they deviate from the other cases, likely due to their low Reynolds number and insufficient separation of scales. The log law intercept CoV values for the proposed transformation in Fig.~\ref{fig:C_Han} and for the GFM transformation in Fig.~\ref{fig:C_GFM} are calculated to be 0.0381 and 0.0549, respectively, representing an approximately 50\% reduction in scatter. The $\kappa$-based CoV values for the proposed and GFM transformations are calculated to be 0.0117 and 0.0559, respectively. It should be noted that, in addition to the reduction in scatter observed for the proposed transformation, the intercept and $\kappa$ values overlap the incompressible range considerably better. \\
\indent  The pre-multiplied mean shear is computed from the new MVT, $\beta_{G}$, and plotted in Fig.~\ref{fig:beta_alpha}. As noted earlier, the slope and intercept were calculated where $d\beta_{G}/dz^*$ is minimum in the log region. This is used as a method to find a region that best approximates a plateau in $\beta$ that is not dependent on assumed bounds of such a region.  Lee \& Moser\cite{LeeMoser15} report that the plateau spans from $z^+$ of 350 to $z/\delta$ = 0.16, and the average value of ${\partial \beta}/{\partial z^+}$ is 6.46e-5 for the LM5200 case. To compare the plateau region across CTBL cases, averages of $\partial \beta_{G}/\partial z^*$ from $z^*$ = 60 $z/\delta$ = 0.15 are calculated. With an exception of M5T1, which does not exhibit a plateau region, potentially due to low Reynolds number, the average of $\partial \beta_{G}/\partial z^*$ for all CTBL cases ranges from 1.16e-3 to 5.87e-4. Thus the average values of $\partial \beta_{G}/\partial z^*$ are comparable to the value calculated from the plateau observed in LM5200. \\
\indent The proposed MVT can also be applied to transform the ITBL cases by setting thermodynamic fluctuation quantities to zero and setting the mean thermodynamic properties to the wall value. The proposed MVT results in mean velocity profiles that collapse the high Reynolds number ITBL cases (LM5200 and BOP4100) with a logarithmic layer starting around $z^*$ of 60 as shown in Fig. \ref{fig:upls_Han}. This logarithmic layer can also be confirmed by the existence of the plateau region in $\beta_{G}$ for LM5200 and BOP4100, as shown in Fig.~\ref{fig:beta_alpha} where the average values of $\partial \beta_{G}/\partial z^+$ from $z^+ = 60$ to $z/\delta= 0.15$ are 2.46e-4 and 2.22e-4, respectively. Despite this success, differences are observed between the incompressible classical log law and the ITBL cases transformed according to the proposed MVT. While the slope of the semi-log region of the MVT-transformed ITBL profiles is very similar to the classical result, differences can be observed in the log layer intercept values (see Fig.~\ref{fig:upls_Han}). The reasons for the differences between the proposed MVT and the classical log-law for incompressible cases is not immediately apparent, however we partially attribute it to observed differences in the value of $\tau_{VG}^+$ in the viscous sublayer for CBTL and IBTL cases, as shown in Fig.~\ref{fig:FVtrad_VS}. Provided that an even more successful scaling of the CTBL and ITBL viscous stress is possible in the viscous sublayer, the proposed MVT might be expected to produce a scaled mean velocity profile for both CTBL and ITBL with the same intercept.\\

\section{Discussion and Conclusion}
In this paper, a generalized total stress-based velocity transformation that takes density and viscosity fluctuations into account has been derived by identifying important characteristics of wall-bounded flows at high Mach and Reynolds numbers. It is demonstrated that the influence of density and viscosity fluctuations are important and must be considered when scaling the turbulent shear stresses of all CTBL cases considered in this paper. When employing Reynolds averaging, fluctuating viscosity or density-related terms exceeded 5\% of the wall shear stress for all cases. In some cases these terms exceeded 20\% of the wall shear stress. Employing the full Favre averaged momentum equation effectively accounts for the influence of density fluctuations on the near wall stress balance but viscosity fluctuation terms were seen to be as large as 12\% of the wall shear stress in the buffer region. \\
\indent When the influences of the density and viscosity fluctuations on the viscous and turbulent stresses are fully accounted for by including all relevant terms in the near-wall momentum equation, two scaling properties have been identified, namely, (1) the Mach-invariance of the near-wall momentum balance for the generalized total stress and (2) the Mach-invariance of the contributions from the generalized viscous and turbulent stresses to the total stress. A new generalized mean velocity transformation, which considers the effects of mean density gradient, both viscous and turbulent stresses, and the effect of density and viscosity fluctuations, has been derived by accounting for these two scaling properties.\\
\indent The proposed velocity transformation is seen to provide an accurate representation of the logarithmic layer. For a wide range of Mach numbers, Reynolds numbers and heat transfer, the scatter in the intercept and slope of the new transformation are within the bounds found for incompressible flows. The new transformation is successful because it accounts for the density and viscosity fluctuation effects in both the viscous and turbulent stresses as well as the relative contributions of the viscous and turbulent stresses to the near-wall momentum balance. For this reason, no Reynolds number dependence was observed in the slope and intercept of transformed velocity profiles under the new transformation. It was successful in collapsing velocity profiles for all compressible cases described in this study.\\
\indent Despite this success, it should be noted that the proposed transformation requires knowledge of the effects of fluctuations in the thermodynamic variables on the viscous and turbulent stresses, which will limit its use in some situations in the near-term. Additionally, the intercept of incompressible flow data when transformed according to the proposed MVT, was found to be shifted relative to the conventional log-law. This is attributed to a small remaining scatter between the incompressible and compressible viscous stress profiles, the source of which is still uncertain. This leads to a shift in the premultiplied shear stress profiles. Further study is needed to explore this effect as both viscous and density fluctuations have been accounted for in the current formulation. \\
\indent While the transformation of Griffin et al.~\cite{GriffinMoin} has shown promising and improved collapse of the mean velocity profiles, the use of the quasi-equilibrium model~\cite{ZhangHussain} to describe the turbulence in the log layer is not strictly accurate, with moderate scatter in the ratio of production and dissipation in the log-layer.
Examination of the underlying assumptions in the derivation of the GFM transformation suggest its remaining challenges may be due to insufficient Mach invariance of the turbulent stresses and the ratio of production and dissipation as traditionally formulated. \\
\indent The success of the proposed MVT can be attributed to considering the influence of density and viscosity fluctuations and the mean property gradients in both the viscous and turbulent stresses. Two scaling properties, namely, the proportionality and the Mach-invariance of the generalized total stress, are identified and integrated into the newly proposed MVT.

\begin{acknowledgments}
This work was supported in part by the Air Force Office of Scientific Research under award number FA9550-20-1-0085 monitored by Dr. Sarah Popkin, and by the National Numerical Evaluation Site for Hypersonics.  Computational resources were provided by the Department of Defense under the High Performance Computing Modernization Program (DoD HPCMP) and by the CRoCCo Laboratory at the University of Maryland.
\end{acknowledgments}

\bibliography{references.bib}

\end{document}